\documentclass[sigchi]{acmart}
\usepackage{booktabs,array}
\newcolumntype{C}{>{\centering\arraybackslash}m{6em}}

\AtBeginDocument{%
  \providecommand\BibTeX{{%
    \normalfont B\kern-0.5em{\scshape i\kern-0.25em b}\kern-0.8em\TeX}}}

\setcopyright{acmcopyright}
\copyrightyear{2020}
\acmYear{2020}
\acmDOI{10.1145/1122445.1122456}

\acmJournal{JACM}
\acmVolume{1}
\acmNumber{1}
\acmArticle{1}
\acmMonth{7}

\newif\ifanderson\andersontrue

\begin{document}

\title{Exploring the Impact of COVID-19 Lockdown on Social Roles and Emotions while Working from Home}
\author{Sam Nolan}
\email{s3723315@student.rmit.edu.au}
\orcid{0000-0001-7451-853X}
\affiliation{%
  \institution{School of Science, RMIT University}
  \streetaddress{GPO Box 2476}
  \city{Melbourne}
  \state{Victoria}
  \country{Australia}
  \postcode{}
}

\author{Shakila Khan Rumi}
\email{shakilakhan.rumi@rmit.edu.au}
\orcid{https://orcid.org/0000-0002-8927-597X}
\affiliation{%
  \institution{School of Science, RMIT University}
  \streetaddress{GPO Box 2476}
  \city{Melbourne}
  \state{Victoria}
  \country{Australia}
  \postcode{}
}

\author{Christoph Anderson}
\email{anderson@uni-kassel.de}
\orcid{0000-0002-4082-8457}

\affiliation{%
  \institution{University of Kassel}
  \streetaddress{Wilhelmsh{\"o}her Allee 73}
  \city{Kassel}
  \state{Hessen}
  \country{Germany}}
  
\author{Klaus David}
\email{david@uni-kassel.de}
\affiliation{%
  \institution{University of Kassel}
  \streetaddress{Wilhelmsh{\"o}her Allee 73}
  \city{Kassel}
  \state{Hessen}
  \postcode{34121}
  \country{Germany}}
  
\author{Flora D. Salim}
\email{flora.salim@rmit.edu.au}
\orcid{0000-0002-1237-1664}

\affiliation{%
  \institution{School of Science, RMIT University}
  \streetaddress{GPO Box 2476}
  \city{Melbourne}
  \state{Victoria}
  \country{Australia}
  \postcode{}
}

\renewcommand{\shortauthors}{S. Nolan, et al.}

\begin{abstract}
  In the opening months of 2020, COVID-19 changed the way for which people work, forcing more people to work from home. This research investigates the impact of COVID-19 on five researchers' work and private roles, happiness, and mobile and desktop activity patterns. Desktop and smartphone application usage were gathered before and during COVID-19. Individuals' roles and happiness were captured through experience sampling. Our analysis show that researchers tend to work more during COVID-19 resulting an imbalance of work and private roles. We also found that as working styles and patterns as well as individual behaviour changed, reported valence distribution was less varied in the later weeks of the pandemic when compared to the start. This shows a resilient adaptation to the disruption caused by the pandemic.
\end{abstract}

\ccsdesc[500]{Human-centered computing~Empirical studies in HCI}

\keywords{behavior logs, productivity, working from home, COVID-19, future of work, affective computing, social roles}

\maketitle

\section{Introduction}
A new coronavirus (SARS-CoV-2) broke out in Wuhan, China in December 2019~\cite{zhou2020clinical}. Within three months of its outbreak, it spread around the world and led the World Health Organization to declare it as global pandemic on the 11th of March~\cite{who_pandemic_declaration}. In Victoria, Australia, the state of emergency was declared on the 16 March 2020~\cite{victoria_state_of_emergency}. The pandemic has changed the way we live our lives, encouraging more people to work from home. This has impacted the way that people work as well as their outlook on the world. Understanding the individual's behaviours during this crisis will help to take steps to mitigate the shock of pandemic on people's lives. Therefore, this paper aims to provide insights on how the COVID-19 lockdown impacted personal lives.

The pandemic offers a unique opportunity to investigate the impact that working from home has on people, and the practical effectiveness of it during and coming out of the pandemic. Although there has been literature on teleworking and the impact of working from home~\cite{dockery_home_family,rupietta_wfh}, the pandemic is a unique situation. The emotions of people in the pandemic need to be considered. This study fills this gap by providing an analysis of the valence and arousal of individuals as they start to work at home. This study analyzes how life has evolved for individuals during this pandemic condition and how their mood and physiological well-being was impacted by it. This study looks into the effects of COVID-19 from three angles. 

Firstly, it contributes to the literature of the psychological effects of COVID-19~\cite{zhang_unprecedented_2020, losada-baltar_were_2020} by looking into the effect that the pandemic has had on the happiness and excitement of individuals before and during the pandemic.

Secondly, it looks into how individuals' social roles have been impacted before and during COVID-19. Social role theory finds that an individual follows different roles or social positions in life, which have different expectations for their own behaviour~\cite{biddle_recent_1986}. Based on the changed working environment due to COVID-19, this study focuses on identifying how often an individual finds themselves in a private or work related role, or a combination of the two. We hypothesised that as the participants starting working from home more, they would combine their work and private roles more often. 

Finally, this paper looks at how valence, arousal, social roles and app activity were impacted by COVID-19. It tries to determine whether the pandemic caused more varied happiness, excitement and computer usage than before COVID-19. And whether that behaviour varies less in the later weeks of lockdown. 
Moreover, we investigate whether individuals' behaviour adapt and change over time to former established routines before COVID-19. 

This paper uses the terms "before the pandemic" and "before COVID-19" interchangeably. These refer to the time before the WHO declared the pandemic on the 11th of March 2020\cite{who_pandemic_declaration}, even though COVID-19 was already in China when we were collecting this data. The "lockdown" is referring to the lockdown in Victoria, Australia, when the state of emergency was declared on the 16th of March. "During the pandemic" refers to any time after the 11th of March.

The research questions this study attempts to provide insight to are
\begin{enumerate}
    \item Whether COVID-19 lockdown has an influence on the distribution of social roles the participants assumed. Particularly, the combining of work and private roles. 
    \item Did the introduction of COVID-19 correlate with a decrease in the happiness distribution of individuals
    \item Did the introduction of COVID-19 correlate with a change the length of the work day?
    \item Were behaviour patterns in valence, arousal, social roles and app activity disrupted during COVID-19? Did they stabilise as the weeks went on?
\end{enumerate}

One of the main challenges with this research is acquisition of similar data for the same individuals that allows for a fair comparison between behaviour before and during the COVID-19 pandemic on a personal level. COVID-19 could not be anticipated and have the data controlled in such a way. To accelerate our understanding to the impact of the pandemic on the individuals, this study uses a novel dataset. The dataset was collected through an application installed on the participants' computer and/or smartphone before and during the lockdown. This allows for a comparison between the behaviour of individuals before and during lockdown. 

In summary the contributions of this research are as follows:
\begin{itemize}
    \item An overview of the effect the lockdown had on the valence distributions of individuals.
    \item A statistical significance test, fisher's exact test is performed to measure the impact of the lockdown on social role of a person.
    \item For the first time identifies and analyses computer and phone behaviour clusters during the pandemic situation. 
    \item An exploration to how much the behaviour patterns of individuals in terms of valence, arousal, social roles and app activity were disrupted during the lockdown.
\end{itemize}

\section{Related work}\label{sec:related}

\subsection{Social Role Analysis}
There has been several investigations into social role theory and it's impact on the way that we interact with computers. People combine their work and private lives in different ways.
Recently, Anderson et al. contributed to this by looking into how social roles can be inferred from ICT usage~\cite{anderson_assessment_2018}. This is moving towards building intelligent attention management systems, controlling when people can be interrupted for tasks that relate to different social roles~\cite{anderson_impact_2019}. 

\subsection{Happiness Measurement}
The literature on COVID-19's impact on valence and individual lives is still developing. Kleinberg et al has recently introduced a new dataset\footnote{\url{https://osf.io/awy7r/} Accessed 3rd of July 2020} collecting emotions of individuals expressed through texts during COVID-19~\cite{kleinberg_measuring_2020}. Furthermore Zhang et al have investigated the impact of COVID-19 in Wuhan, and call to pay attention to the mental and physical impacts of those not affected by the virus epidemiological~\cite{zhang_unprecedented_2020}.

Another source of COVID-19's impact on individual happiness comes from the Hedonometer. The Hedonometer is a way of attempting to measure the general happiness of the population by observing the word usage of tweets~\cite{dodds_temporal_2011}.  This project found that although the COVID-19 caused a large dip in happiness, over time happiness has rebounded to be close to levels before COVID-19, asides from the dip in happiness caused by the murder of George Floyd and the consequent protests against police brutality. \footnote{\url{https://hedonometer.org/timeseries/en_all/} Accessed on the 1st of July 2020}. This would hypothesise that although a dip in initial happiness may be observed, over time, the impact on happiness would be negligible.

\subsection{App Usage Behaviour}
Many of the existing literature regarding app usage behaviour analysis focused on how people use the apps in their smartphones~\cite{ferreira2014contextual,verkasalo2009contextual}. These research suggested that mobile usage pattern of a person contain many micro-usage e.g, locking or unlocking phones. Some researchers worked towards predicting the next app which is more likely to be opened by the user~\cite{huang2012predicting,liao2013feature}. Jones et al. tried to determine which smartphone apps encourage mobile phone usage behaviour. The have analysed the revisiting patterns of different apps for 165 users. They found that the app usage behaviour on macro-level is technology independent~\cite{jones2015revisitation}. 

\subsection{Working from home}
The effects of working from home and work-related mobile phone usage in the after-hours have been investigated in the field of organizational psychology and work management. On the one hand, positive effects on individuals' well-being have been found when, for example, work-related usage of mobile phones is in line with their motivation and individual preferences\cite{Ohly2014}. Also, the possibility to meet demands from both domains -- work and private -- have been shown to have positive effects on well-being, especially for individuals who like to integrate work and private-related matters\cite{Derks2016}. On the other hand, adverse effects related to conflicts between work and private domains\cite{Delanoeije2019} as well as on individuals' well-being have been reported -- originating from the blurring of \textit{work} and \textit{non-work} domains\cite{Mellner2016}.


\subsection{Remarks}
Due to the pandemic, a large number of people are currently working from home which has impact on daily app usage patterns. The social role and temporal patterns in app use are different than previous times. Being attentive to the feelings plays vital role in staying healthy during this tough time. This study shed light on all of these by analyzing the data during COVID-19. 
\section{Dataset Description}\label{sec:data}
Among others, the dataset comprises information on application usage, locations, social roles and valence and arousal ratings from five participants. All the participants in the dataset are information workers in Victoria, Australia. The data was collected using an application that was installed on their computer and/or their smartphone. The application was developed by Uni Kassel, Germany. All data was gathered with prior written permission from participants. 
The first data collection took place from the 20th of January to the 26th of February 2020. The second data collection started during COVID-19 -- 17th of March to the 21st of May 2020. In the following, we present details on the collected data.

\subsection{ESM surveys}\label{sec:esm}
This application used Experience Sampling Method ESM and gave a survey for the user to complete every 90 minutes that the user was active on their devices. This survey asked:
\begin{itemize}
    \item How happy were you feeling in the last hour? (Valence: Scale from 1 to 5)
    \item How excited were you feeling in the last hour? (Arousal: Scale from 1 to 5)
    \item What role you were fulfilling in the past 15 minutes? (Work, Private, or Both)
\end{itemize}

These surveys were optional and were collected before the outbreak of COVID-19 and during the outbreak. A total of 488 and 436 surveys were gathered before and during COVID-19, respectively.

\begin{table}[ht]
	\centering
	\caption{Quantity of survey collected before and during COVID-19 from each participant}
	\begin{tabular}{lcc}
		\toprule
		              & before & during \\
		\midrule
		Participant 1 & 124    & 105    \\
		Participant 2 & 217    & 9      \\
		Participant 3 & 107    & 55     \\
		Participant 4 & 0 &    222 \\
		Participant 5 & 0 & 45 \\
		Total         & 488    & 436    \\
		\bottomrule
	\end{tabular}
	\label{tab:eventcollection}
\end{table} 

\begin{figure}[ht]
    \centering
    \includegraphics[width=0.7\linewidth]{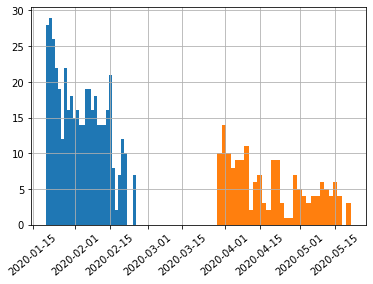}
    \caption{Histogram of surveys collected before and during COVID-19}
    \label{fig:eventcollection}
\end{figure}

For the comparison in happiness and roles, only the surveys of those who participated during and before the pandemic were used. The number of survey responses that were before and during COVID-19 from each participant is noted in Table~\ref{tab:eventcollection}. The histogram of surveys collected during COVID-19 is illustrated in Figure~\ref{fig:eventcollection}. The left group is the surveys collected before COVID-19 and the right the surveys collected during.

\subsection{Application usage}\label{sec:appusage}
The application also recorded the application usage of the user. Recording the application that is currently active and the time it was opened for both smartphone and desktop.

This study filtered out windows utilities, (Desktop, start menu etc), Android home screen, and applications that were only used very little to make a dataset of over 174,000 records of usage of different applications over the time measurement period. The sample size used for the analysis is noted in Table~\ref{tab:appeventcollection}.
\begin{table}[ht]
    \centering
    \caption{Samples used for the analysis of application usage before and during COVID-19 for all 5 participants}
    \begin{tabular}{lcc}
    \toprule
    Participant & before & during \\
    \midrule
    Participant 1 & 18037 & 30326 \\
    Participant 2 & 35566 & 5431 \\
    Participant 3 & 17330 & 18671 \\
    Participant 4 & 0 & 25472 \\
    Participant 5 & 0  & 23933 \\
    Total & 70933 & 103833 \\
    \bottomrule
    \end{tabular}
    \label{tab:appeventcollection}
\end{table}
These applications were then categorised into 12 categories. 
The categories and examples of applications are below:
\begin{itemize}
    \item Browser (Chrome, Microsoft Edge)
    \item Communication (Discord, Messenger, WeChat, Zoom)
    \item Developer Tools (Command line, Python, Sublime Text)
    \item Entertainment (YouTube, Google Play Music, Spotify)
    \item Game (Steam and other Games)
    \item Media Viewing (VLC, Windows Media Player, Android Photos)
    \item Reading (PDF viewers, Audible, Wuxia World)
    \item Settings (Device maintenance, Android Settings)
    \item Tools (Google Drive, ING Banking, tramTRACKER)
    \item Work Communication (Email tools, Teams)
    \item Utilities (Snipping Tool, Camera, Contacts, Phone)
    \item Writing (TeX Studio, Notepad, One Note, Microsoft PowerPoint)
\end{itemize}

\section{Methods}\label{method}
Our methods involved first looking at an overall comparison between the survey responses before and during COVID-19 in section~\ref{methodesm} and move into analysis of how COVID-19 disrupted valence, arousal, social roles and app usage.

\subsection{ESM Analysis}\label{methodesm}
This study first compares the roles assumed by the participants. The aim of this is two fold. Firstly, to determine whether the pandemic introduced a change in social roles. Secondly, whether the introduction of COVID-19 caused a caused the proportion of work roles chosen each day to change.

We used \textit{Fisher's exact test} for the first comparison. \textit{Fisher's exact test} is a statistical method of determining whether two variables are independent or dependent of each other. It is particularly appropriate when the number of samples in the distribution is small\cite{campbell2007chi}, which is true in our case and particularly for participant 2. The null hypothesis of the test is that the two distributions are independent.

The \textit{Fisher's exact test} used to compare the dependence of two variables, the amount a person chose a role, and whether or not the participant was taking the surveys during the lockdown. Below is a sample table shown for participant 1 for which \textit{Fisher's exact test} was run.

\begin{table}[h]
    \centering
    \caption{Sample Fisher's exact table used for input in participant 1}
    \begin{tabular}{rcccc}
    \toprule
         & Private &  Work and Private & Work \\
    \midrule
before frequencies & 51 & 22 & 51  \\
during frequencies & 32 & 21 & 52  \\
    \bottomrule
\end{tabular}

    \label{tab:sample_fisher}
\end{table}

To determine whether the amount of work done in a day changed, we found the proportion that a user would choose a work role (that is, Work or Work and Private). Then did a \textit{t-test} to determine whether these proportions changed before and during COVID-19. Box plots were also used to show the distribution of data.

To determine if the mean of the valence distribution changed before and during COVID-19, a \textit{t-test}, which is a test to determine whether the mean of two distributions are different, was deployed. This test is to determine whether the average valence of each participant before and after COVID-19 changed.

\subsection{Stability of features in COVID-19}

One of our research question was to determine whether before COVID-19, the app usage activity, happiness, excitability and social role would be more consistent than the earlier days of COVID-19 and whether the early days of COVID-19 would be more varied than the later days of COVID-19.

This study takes inspiration from Sadri et al\cite{sadri_rest_of_day} who tried to predict the afternoon movements of individuals based off what they did in the morning. They did this by determining how similar each day was to all other days within the observed window of time. A similar approach was taken in looking at how similar each day was to each other day in different time periods.

This study investigates the impact of the COVID-19 lockdown on these four variables. The first three were gathered in the ESM surveys presented to the users:

First, valence. Valence was gathered by asking the happiness of the individual. The valence is measured between 1-5.

Secondly, arousal. Arousal was gathered by asking participants how excited they feel, and is also measured between 1-5.

Thirdly, social role. The social role was a categorical value of "work" "private" or "work and private"

Finally, app usage. The app usage was correlated by segmenting the data with IGTS (described below)

\begin{figure}[ht]
    \centering
    \includegraphics[width=0.7\linewidth]{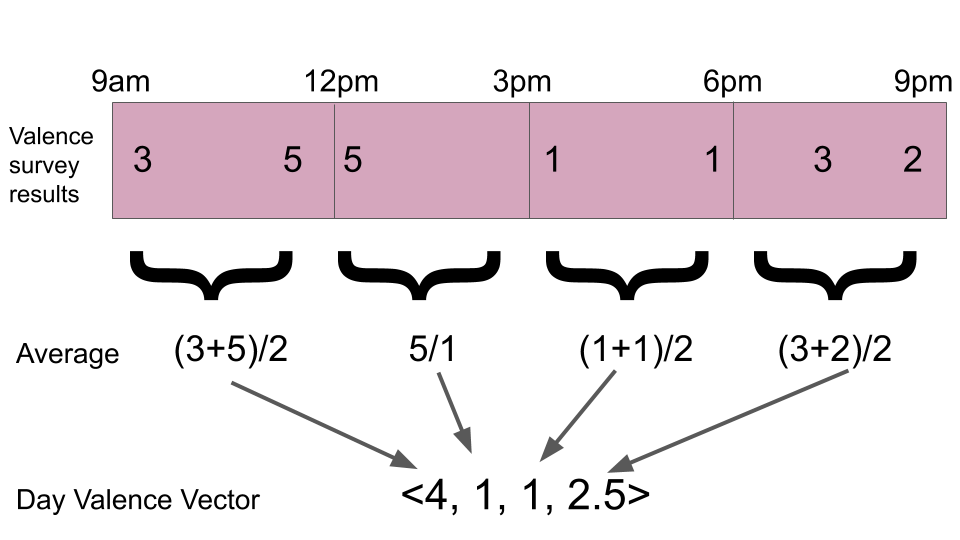}
    \caption{An example of transforming a day of valence survey results to a vector}
    \label{fig:vectorprocess}
    \Description{The process is to first find the average happiness for 3 hour windows between 9am and 9pm. Then these 4 averages are turned into a vector}
\end{figure}

To determine how the valence, arousal and social role of a participant changed over the days before and during COVID-19, a four dimensional vector was created for each day, where each element was the average value recorded in the surveys between hours $[9-12)$,  $[12-15)$, $[15-18)$, $[21-24)$ respectively, four three-hour segments. We chose these time intervals as these were the times that our participants were working.

The similarity matrix between each day for each of the ESM survey vectors was measured using \textit{Jaccard index}, which is a popular method of similarity measurement between two finite sets specifically for sets categorical variables~\cite{niwattanakul2013using}. It measures the similarity between two sets by dividing the common number of features with total number of available features. \textit{Jaccard distance} is the inverse of jaccard similarity which can be obtained by using the equation~\ref{Jac:Eq}.
\begin{equation}
Jaccard'Dist(A,B)=1-\frac{|A \cap B|}{|A \cup B|}
    \label{Jac:Eq}
\end{equation}
As social role is a categorical variable, we compute Jaccard distance to find the non-similar days of a user during different periods of interest (before, 2 weeks after and 6 weeks after the pandemic declaration).

\label{appusageanalysis}
\begin{figure}[ht]
    \centering
    \includegraphics[width=\linewidth]{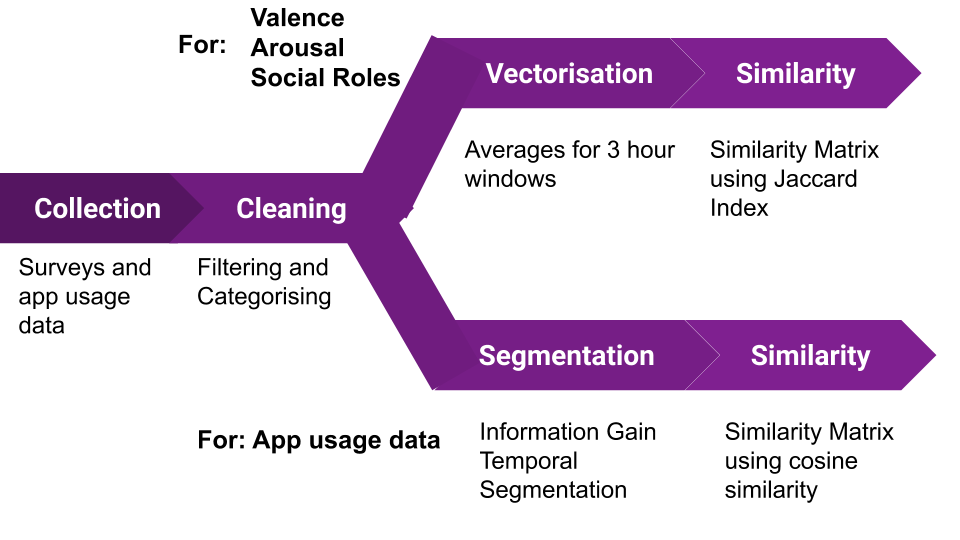}
    \caption{Process for ESM survey and Application usage analysis}
    \label{fig:process}
    \Description{The process used for this study was first, data collection. Then cleaning (which included removing little used applications) and categorising. Then Segmentation using IGTS with included finding an appropriate k using a knee point function. Then finding the similarity between the clusters of each day and presenting it}
\end{figure}

\begin{figure}[ht]
    \centering
    \includegraphics[width=\linewidth]{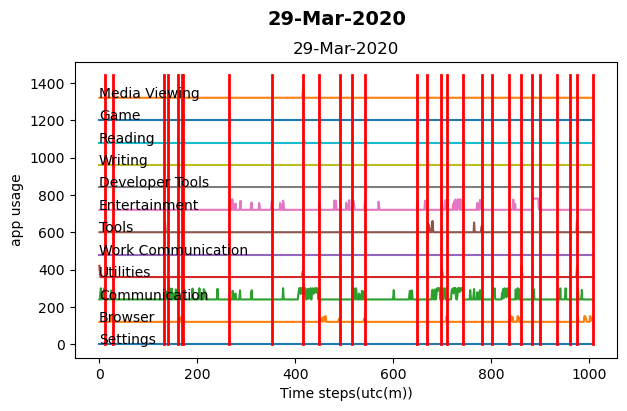}
    \caption{An example segmentation of a day's app usage}
    \label{fig:examplesegmentation}
    \Description{There are 12 lines that run from left to right. The lines goes up and down to indicate an app of that category is being used. There are 28 vertical red lines indicating the segments that were found in the time series}
\end{figure}

\textit{Information Gain Temporal Segmentation}~\cite{sadri_information_2017} (IGTS) was applied to segment the app usage time series into different activities that the participant did throughout their day. IGTS is a method of segmenting multivariate times series by attempting to reduce the amount of entropy in each individual segment. The method creates segments that in total have the lowest entropy. IGTS was applied to each day to discern the activities that were completed throughout the day. To use IGTS, an appropriate value for the parameter $k$ must be chosen. $k$ is the amount of splits to be found in each day (the amount of segments is therefore $k+1$). Including more segments increases information gain. Information gain is the amount of entropy lost by segmenting the time series. However, too many segments decreases the size and interpretability of the segments found. So choosing a $k$ that finds a good balance between the interpretability and size of clusters is important. The original IGTS paper offers a method of finding the knee point in the information gain vs $k$ curve. This method was used over all the segments for all users for all days and the median $k$ was chosen for this analysis. The median $k$ was found to be 27. An example segmentation for a day is in figure~\ref{fig:examplesegmentation}.

IGTS was then applied over all the time series, giving a list of segments for each user.

We compare the similarities between days in terms of app usage using cosine similarity. Cosine similarity is measured based on the dot product of two vectors. Thus, the angle between two vectors is considered in measuring the similarity. Given two $N$ dimension vectors $\vec{A}$ and $\vec{B}$, the cosine similarity is computed using the equation~\ref{eq:cos_sim}.
\begin{equation}
Cosine(\vec{A},\vec{B})=\frac{\vec{A} \cdot \vec{B} }{|\vec{A}| |\vec{B}|}
    \label{eq:cos_sim}
\end{equation}

In our experiment, the vector of each day contains the information gain of each type of app usage in a segment. We considered maximum activity of an app category in a segment. Considering total 12 app categories and 28 segments, each vector contains 336 elements. We applied the pairwise cosine similarity on the vectors corresponding to each day to identify the similar days of each user based on app usage activity during different periods.  
\section{Results}\label{results}

\subsection{Work and Private Roles}\label{workprivate}
\begin{figure}[ht]
    \centering
    \includegraphics[width=0.7\linewidth]{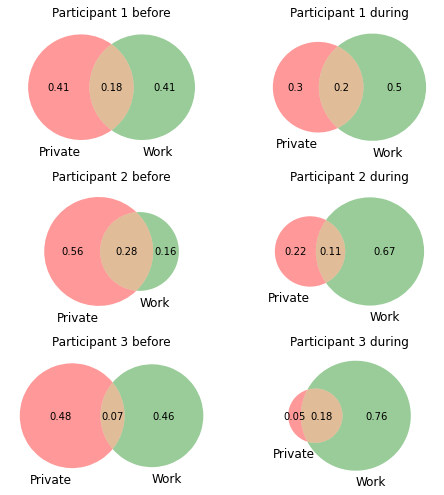}
    \caption{Combination of Work and Private roles before and during COVID-19 pandemic. The rows indicate participants and the left column is before COVID 19 and the right column is after.}
    \label{fig:venn}
    \Description{Six venn diagrams showing the frequency of work and private roles, with 2 columns indicating before COVID-19 and during COVID-19, and three rows for each of the three participants. Only the last 2 show a significant difference between work and life roles. The second venn diagram for during only has a total of 9 responses.}
\end{figure}

The venn diagram in Figure~\ref{fig:venn} shows the proportions for which participants found themselves in different roles. Participant 2 and 3 showed a large increase in the proportion of work roles assumed during the lockdown.
It also shows the heavily skewed work role being assumed during the lockdown period in comparison to private only and both work and private roles.

To determine whether the results shown in the Figure~\ref{fig:venn} were significant, we used fisher's exact test twice.  The null hypothesis for the first test is that the roles that an individual chose were independent of whether COVID-19 is currently occurring. The $p$ values for this test are reported in Table~\ref{tab:fisherexactfull}. The null hypothesis for the second test was that choosing a combined role over a separate role was independent of the pandemic. The $p$ values are reported in Table~\ref{tab:fisherexact}.

\begin{table}[ht]
    \centering
    \caption{Results for a fisher's exact test of independence for social roles and the lockdown}
    \begin{tabular}{rc}
    \toprule
  Participant & $p$ value \\
    \midrule
    Participant 1 & 0.2434 \\
Participant 2 &	0.002156 \\
Participant 3 &	$1.877\cdot 10^{-8}$ 	\\
   \bottomrule
   \end{tabular}

    \label{tab:fisherexactfull}
\end{table}

\begin{table}[ht]
    \centering
    \caption{Results for a fisher's exact test of independence on Work and COVID-19}
    \begin{tabular}{rcc}
    \toprule
  Participant & Fisher exact Statistic & $p$ value \\
    \midrule
    Participant 1 &	1.159091  & 	0.734999 \\
Participant 2 &	0.319672 &	0.450184 \\
Participant 3 &	3.174603 &	0.030161 \\
  \bottomrule
\end{tabular}

    \label{tab:fisherexact}
\end{table}

As shown in Table~\ref{tab:fisherexactfull}, the pandemic had a significant impact on the social roles specifically for participants 2 and 3. In Table~\ref{tab:fisherexact}, we observe that only participant 3 had combined social roles significantly.

\subsection{Longer work hours daily}
\begin{figure}[ht]
    \centering
    \includegraphics[width=0.7\linewidth]{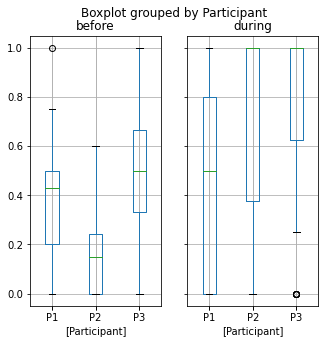}
    \caption{Proportions of the day reported in a work state, before and during COVID-19}
    \label{fig:workdaybox}
    \Description{Six boxplots in two boxes, one before and the other during COVID-19. The boxplots show that all participants increased the amount of work that they did during COVID-19}
\end{figure}

To investigate whether COVID-19 caused a change in the size of proportion of work related roles assumed in a day, we charted the proportions of work assumed by each participant in a box plot shown in Figure~\ref{fig:workdaybox}. The significance of these results were determined using t-test and are found in Table~\ref{tab:workdays}.

\begin{table}[ht]
    \centering
    \caption{Results for a t-test on the change of work day size}
    \begin{tabular}{rcc}
    \toprule
  Participant & test statistic & $p$ value \\
    \midrule
    P1 & -1.318339  & 0.191814\\
    P2 & -4.874137 	& 0.000022\\
    P3 & 	-3.211650 & 	0.002138\\
  \bottomrule
\end{tabular}

    \label{tab:workdays}
\end{table}

This shows that 2 out of 3 participants showed a significant change in work, increasing the amount of work in a day during COVID-19, and all participants increased the proportion of work in their days. This confirms our hypothesis that the lockdown has an influence on longer hours dedicated to work daily. 

\subsection{Valence}\label{valence}
Next, We analysed whether the valence distribution significantly changed during  COVID-19 compare to before  COVID-19 or not. A t-test of independence was performed to compare the distribution before and during COVID-19. The null hypothesis of the test was that the distributions of scores given (before and during COVID-19) have the same mean.

In Table~\ref{tab:hapstat}, $\mu$ is the sample mean before COVID. $\sigma$ is the sample standard deviation before COVID-19. $\mu'$ is the sample mean during COVID-19 and $\sigma'$ is the standard deviation during COVID-19.

\begin{table}
    \caption{Happiness Statistics for Participants}
    \label{tab:hapstat}
    \begin{tabular}{rcccccc}
    \toprule
    Participant & $\mu$ & $\sigma$ & $\mu'$ & $\sigma'$ & t-test statistic & p value \\
    \midrule
    P1 & 2.95 &	0.54 &	2.91 &	0.48 &	0.56 & 0.578 \\
    P2 & 3.34 &	0.82 &	3.44 &	0.88 &	-0.38 & 0.70 \\
    P3 & 3.21 &	0.66 &  2.90 & 	0.48 & 	2.97 & 0.0034 \\
    \bottomrule
    \end{tabular}
\end{table}

It was found that only one participant (again, Participant 3) had a significant difference in reported happiness. This also shows that for the majority of participants, there was not a significant change in valence. This does not provide strong evidence that valence was impacted by the lockdown and working from home. This, however, only shows an aggregate analysis. Later, in section \ref{covidimpact} we present an analysis of the fine-grained changes in the distribution of valence during lockdown.

\begin{figure}[ht]
    \centering
     \includegraphics[width=\linewidth]{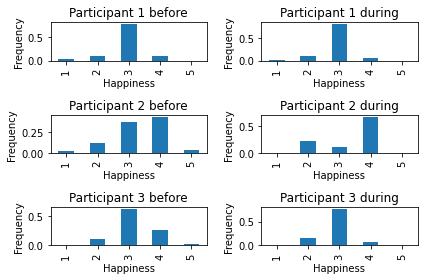}
    \caption{Bar charts of happiness distributions. Left column is before COVID, Right column is after. Rows are participants}
    \label{fig:happyhist}
    \Description{Bar charts of Happiness distributions. These distributions show little change before and during COVID-19}
\end{figure}

Figure~\ref{fig:happyhist} shows the distributions of reported happiness for each participant. Participant 1 and 3 are similar. Participant 2 shows a difference in distribution, however, it is not significant because Participant 2 did not have a large sample size of surveys during the pandemic.

\subsection{COVID-19 lockdown's impact on behaviour}\label{covidimpact}

The next research question is how COVID-19 impacted the stability of valence, arousal, social role and app usage. We are reporting the behaviour distribution similarity for participant 1 (with his associated desktop app usage) and participant 4 (with his associated mobile app usage).

\begin{figure*}[t]
    \centering
    \includegraphics[width=0.24\linewidth]{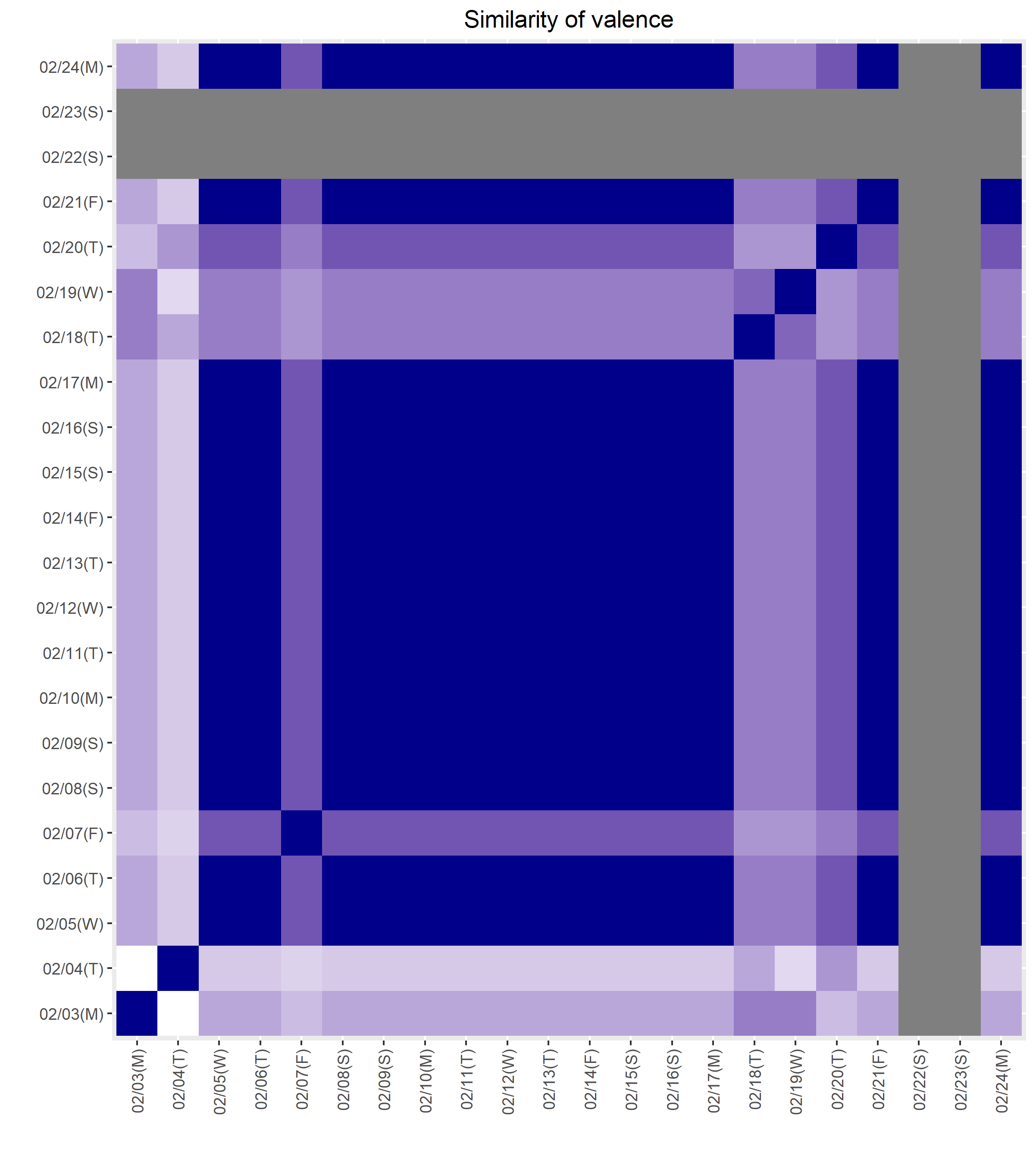}
    \includegraphics[width=0.24\linewidth]{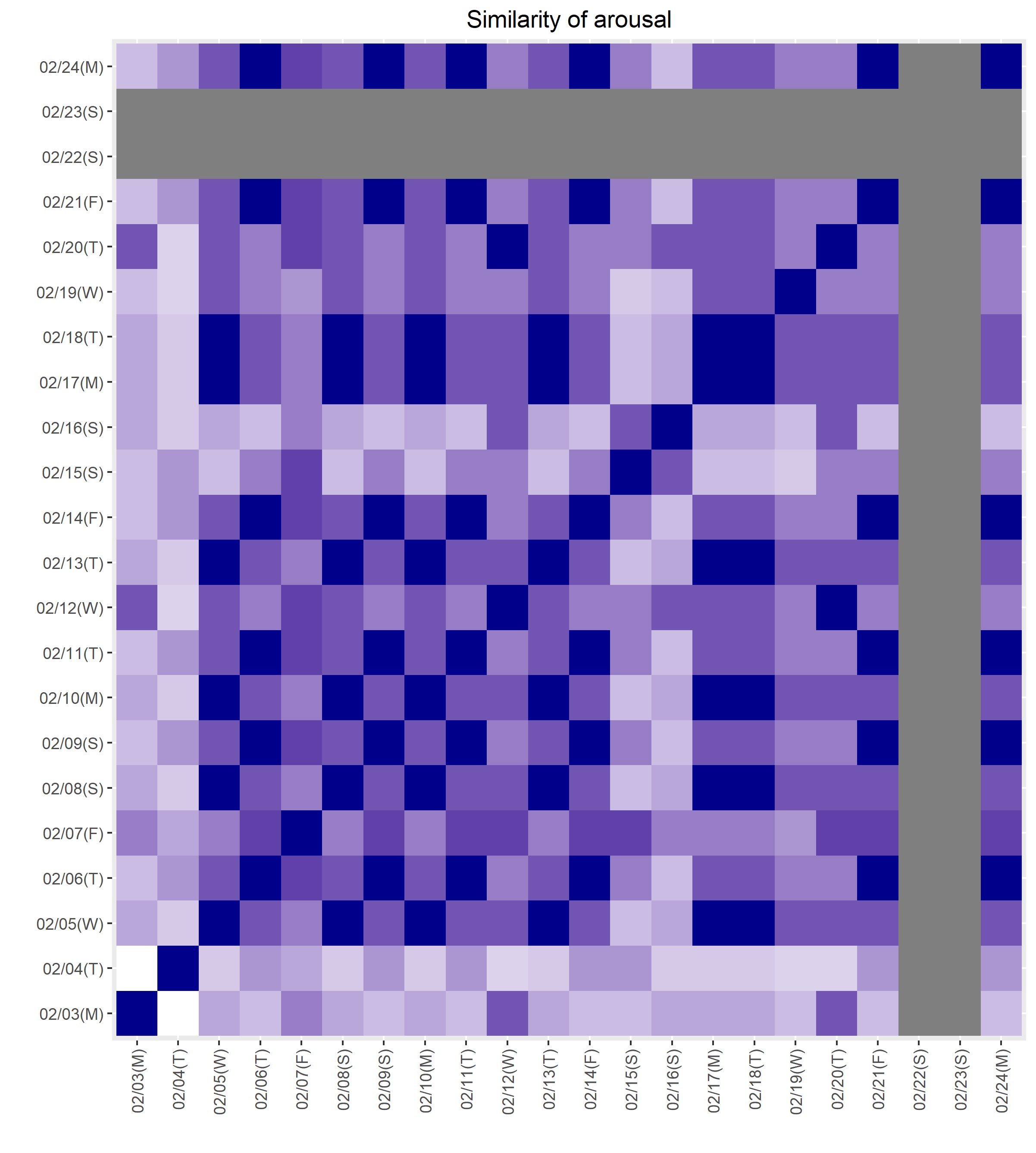}
    \includegraphics[width=0.24\linewidth]{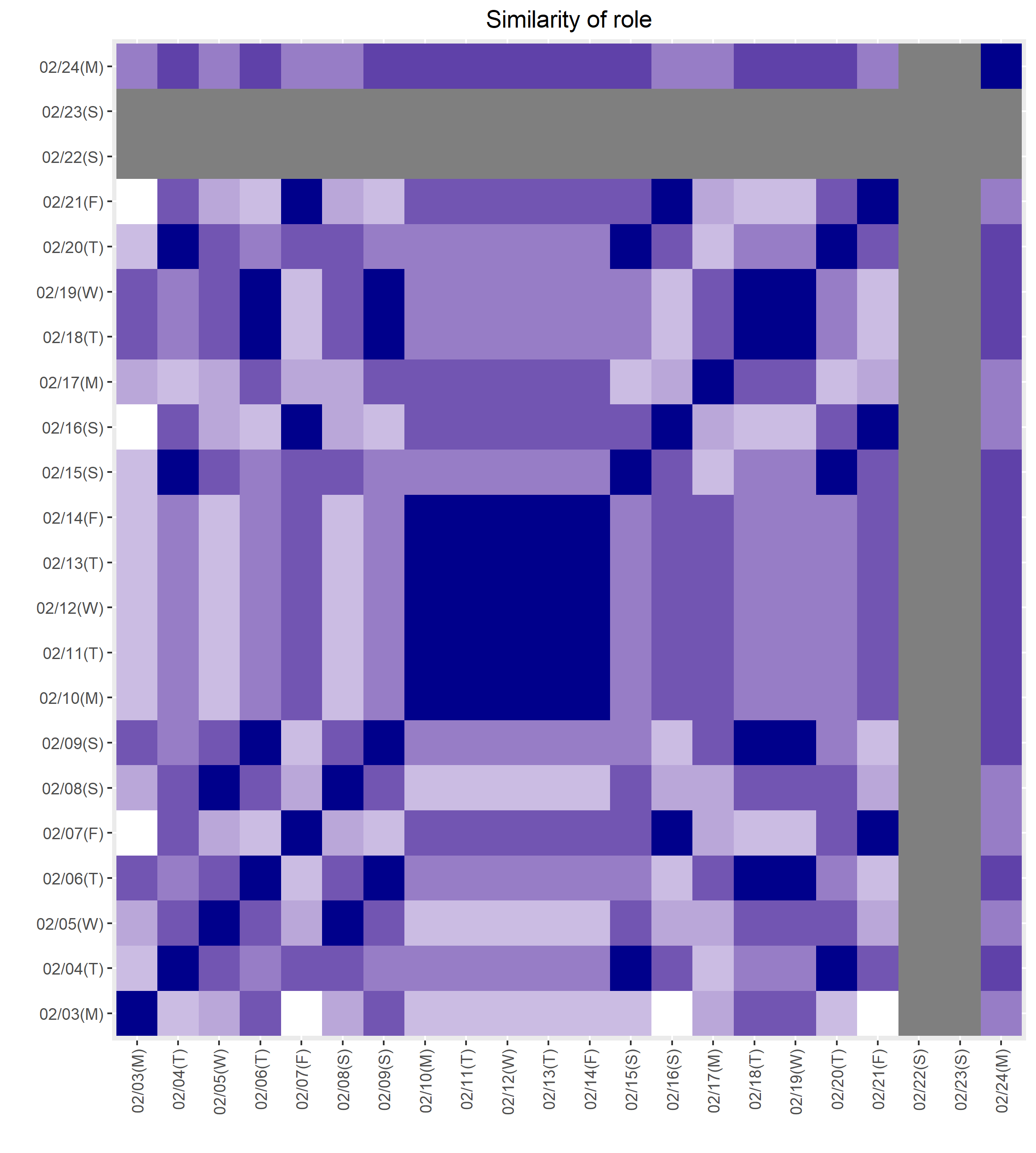}
    \includegraphics[width=0.24\linewidth]{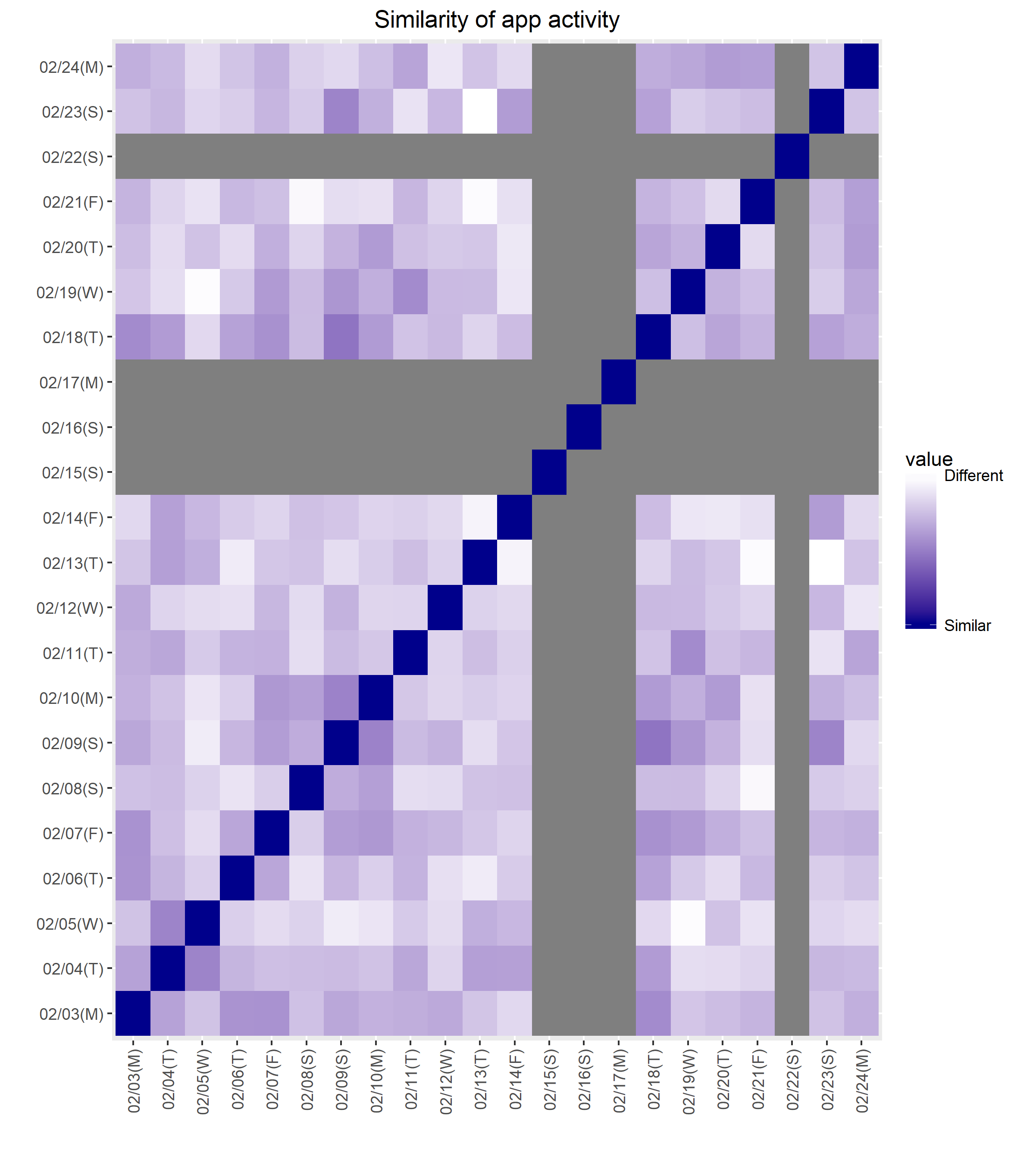}
    
    \includegraphics[width=0.24\linewidth]{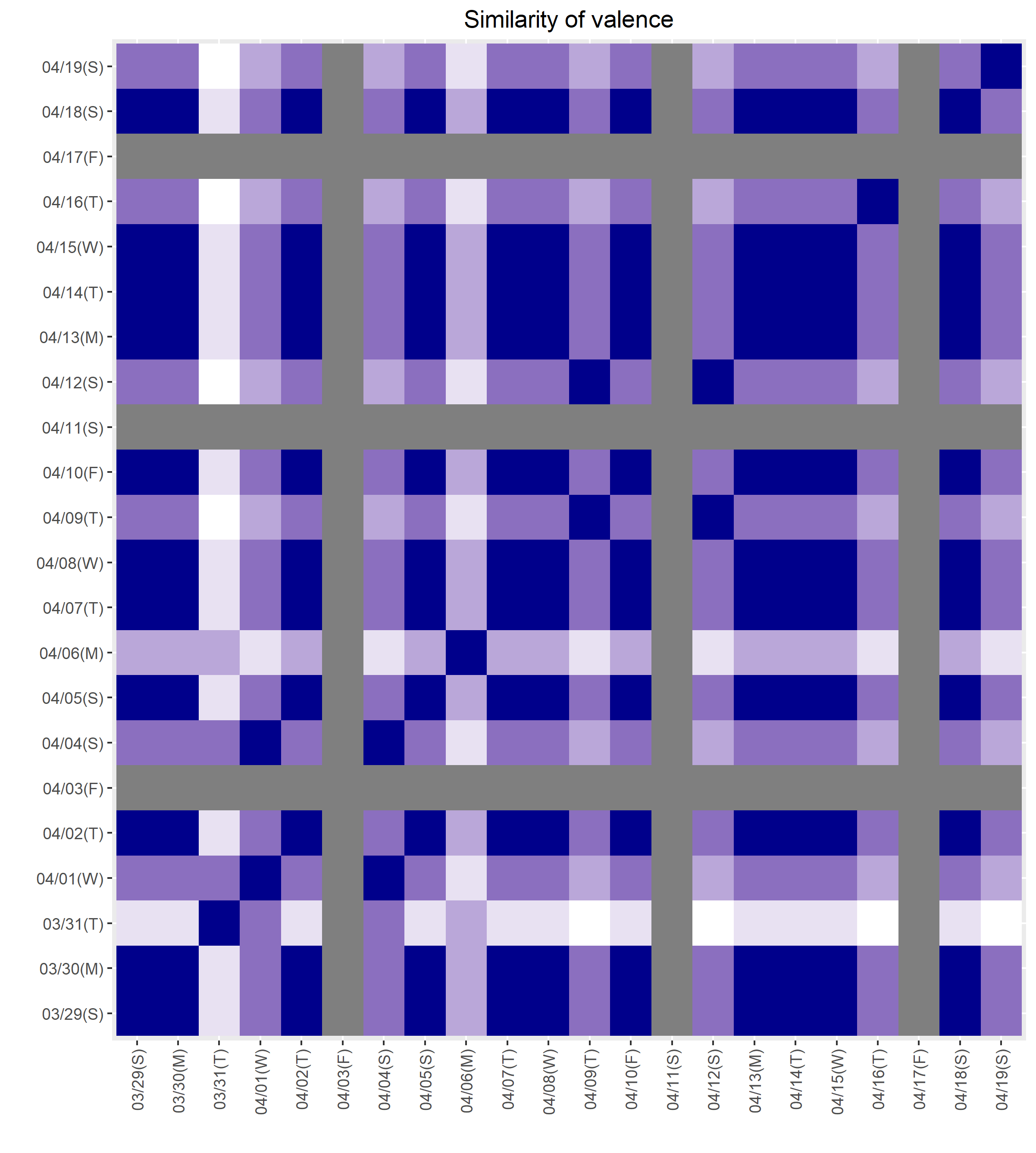} 
    \includegraphics[width=0.24\linewidth]{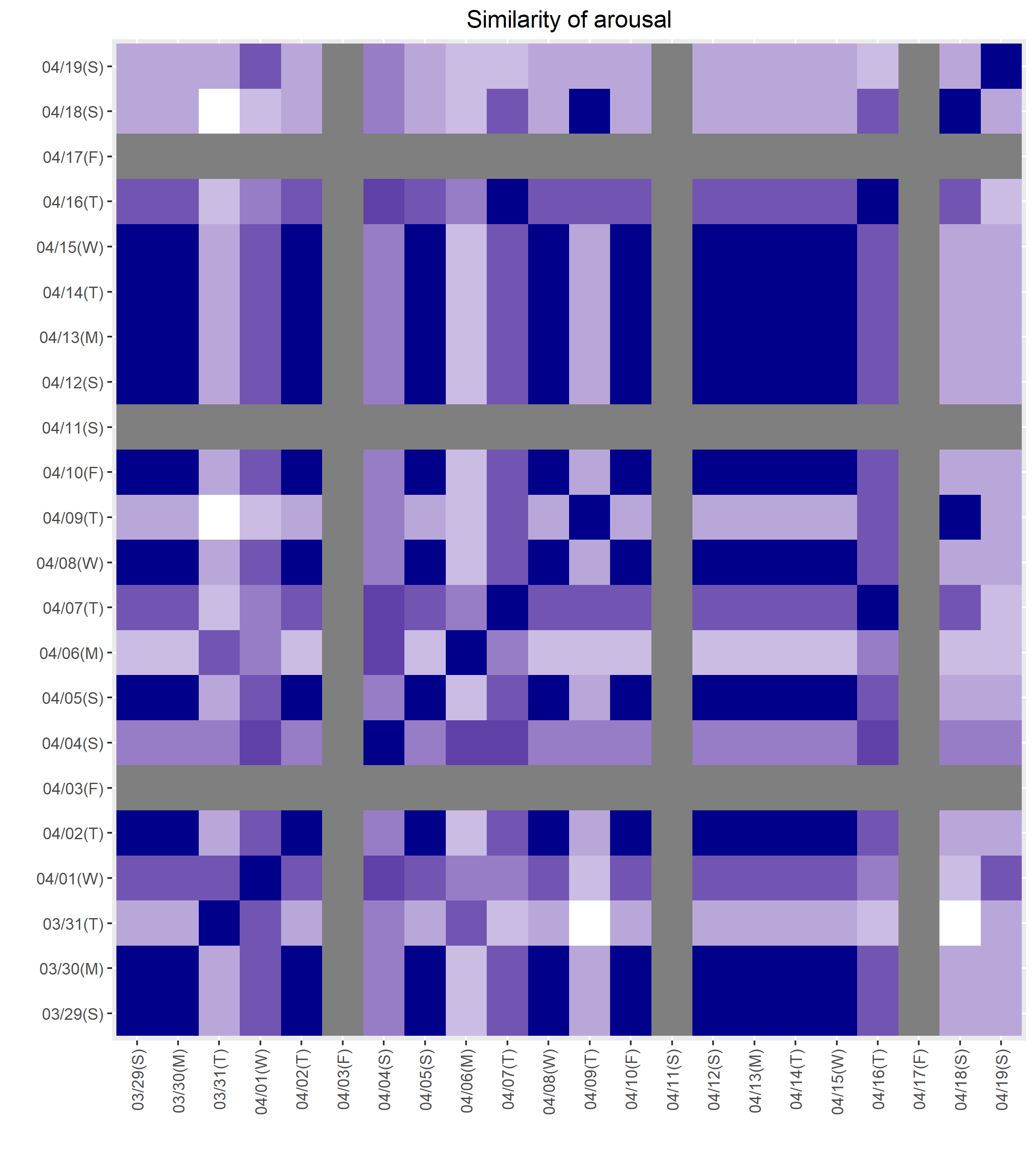}
    \includegraphics[width=0.24\linewidth]{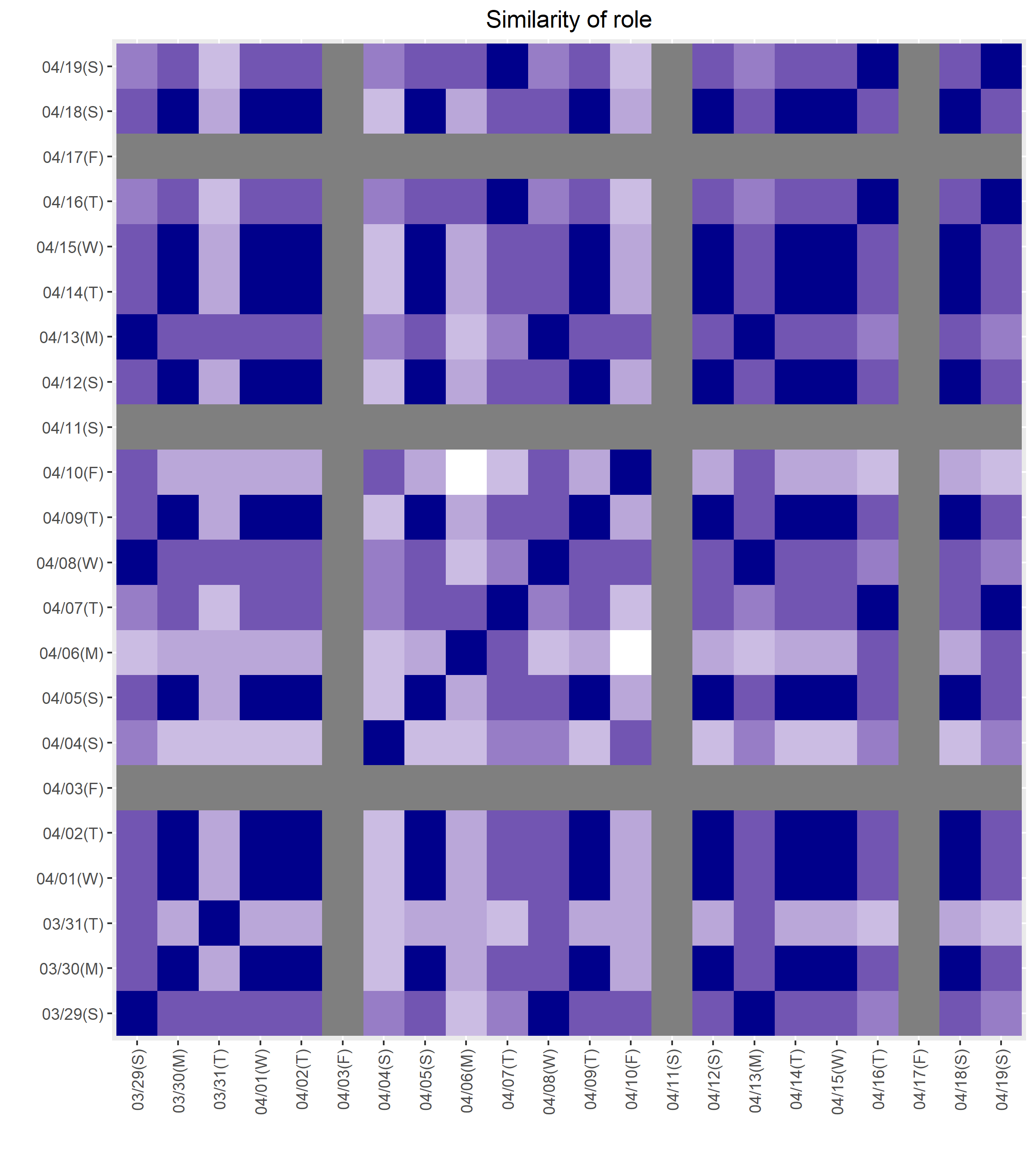}
    \includegraphics[width=0.24\linewidth]{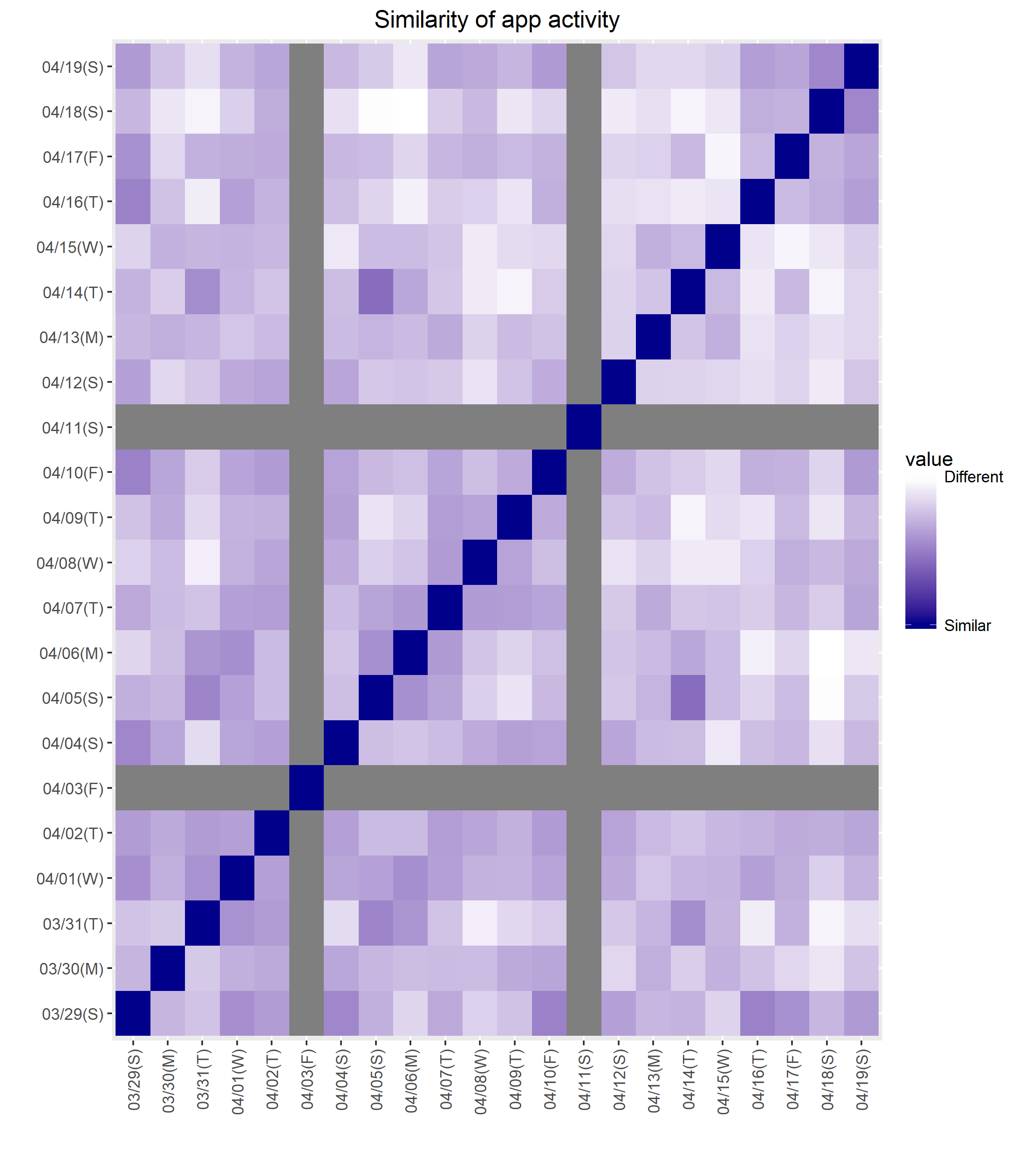}
    
    \includegraphics[width=0.24\linewidth]{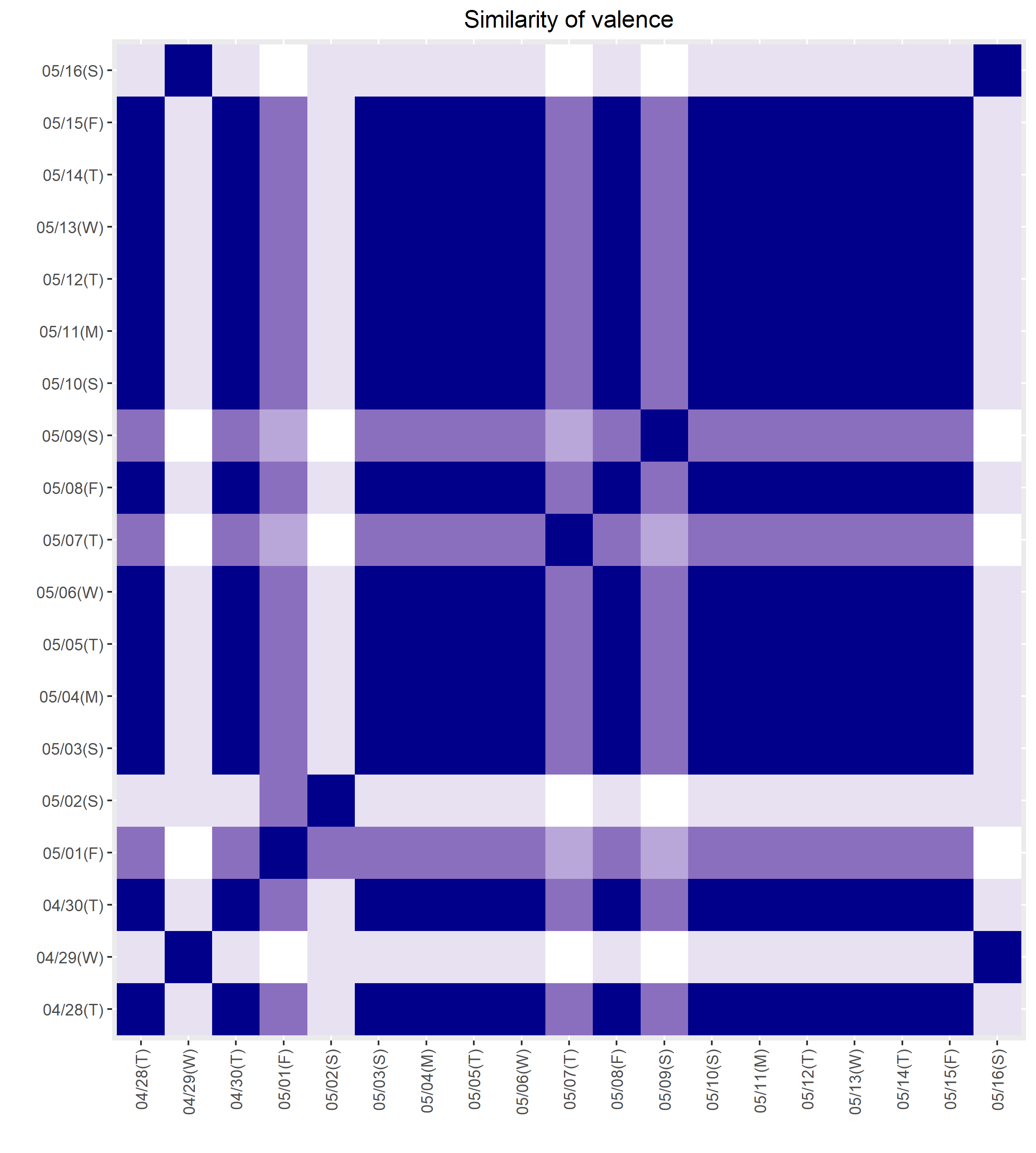}
    \includegraphics[width=0.24\linewidth]{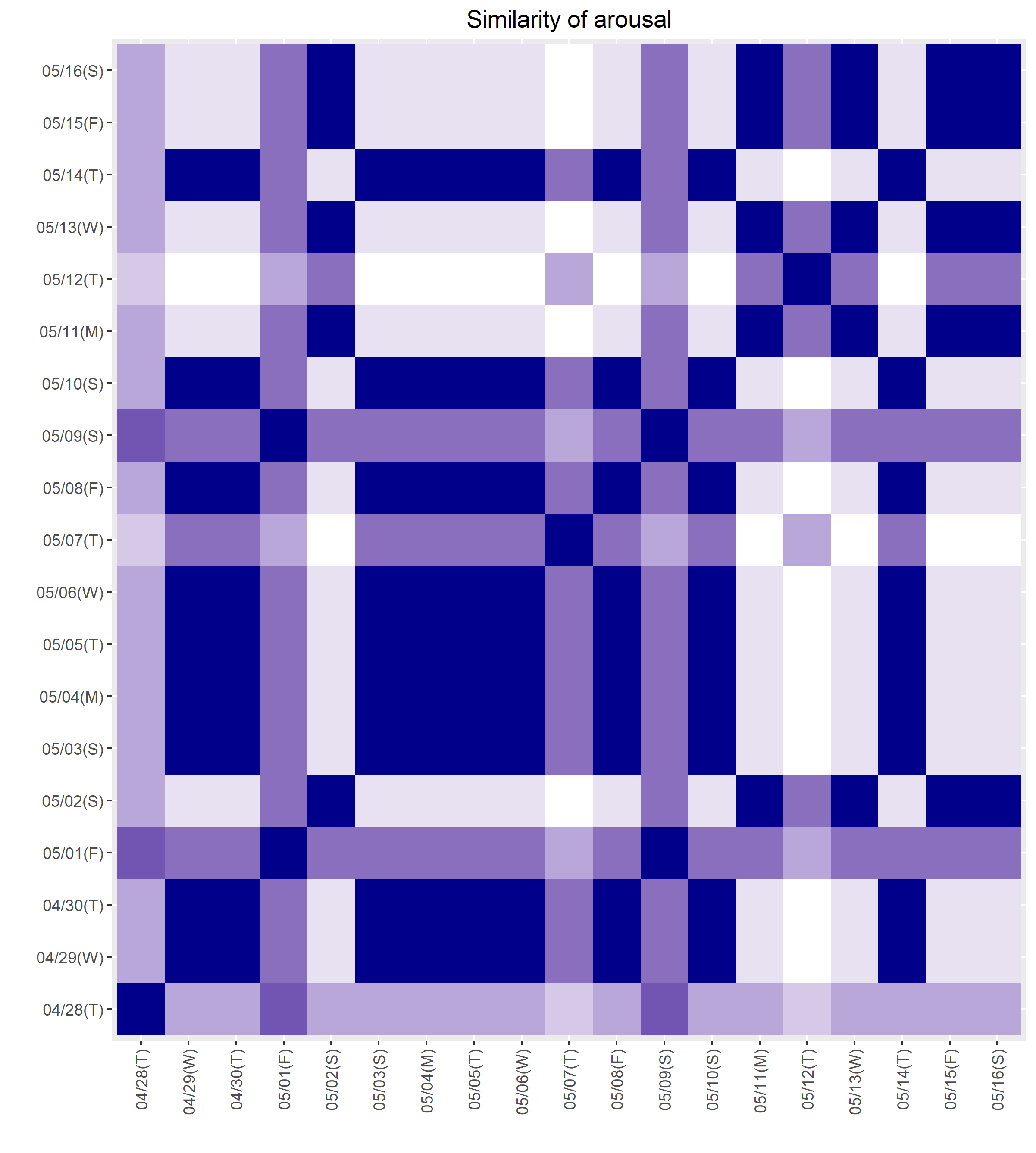}
    \includegraphics[width=0.24\linewidth]{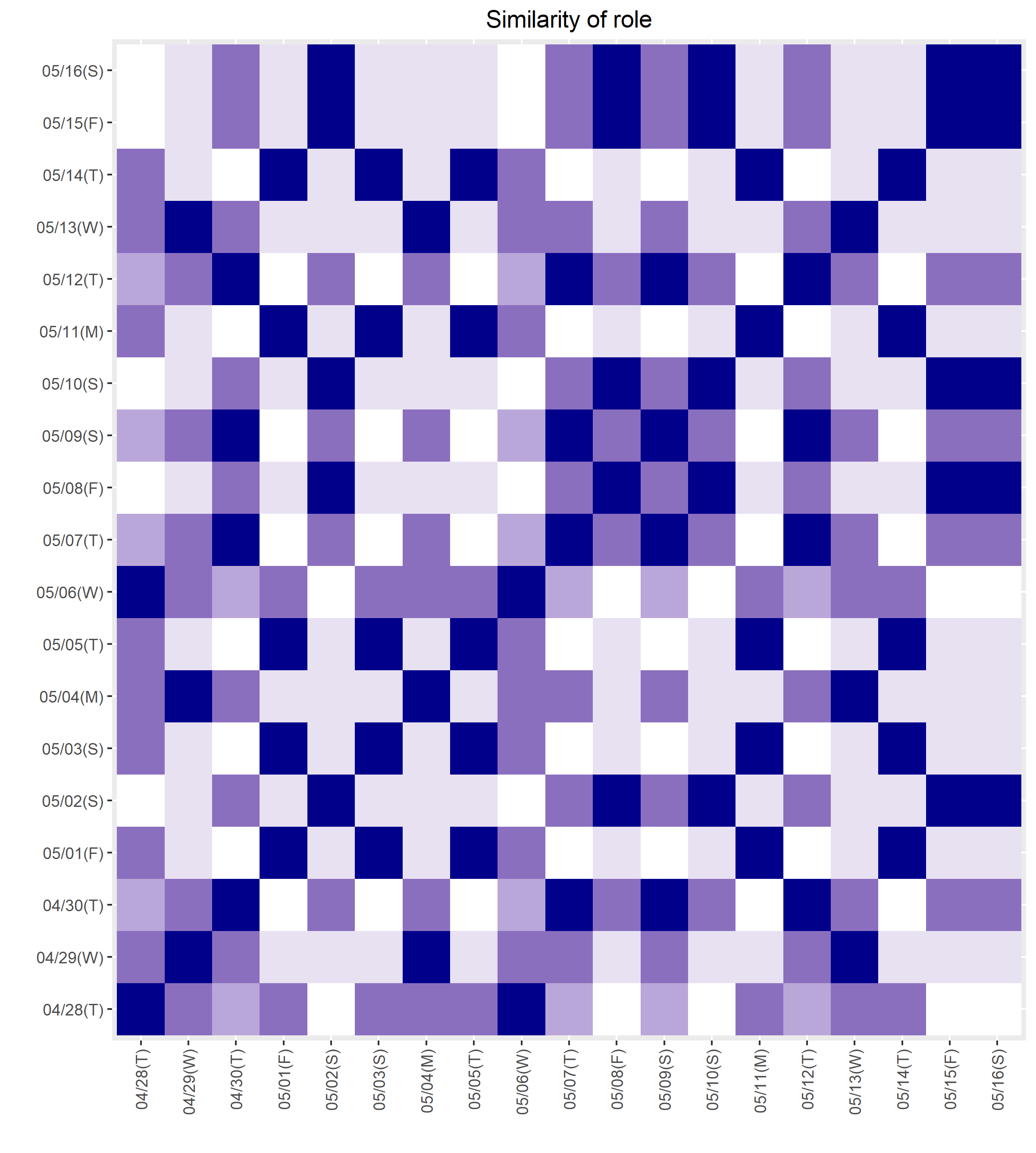}
    \includegraphics[width=0.24\linewidth]{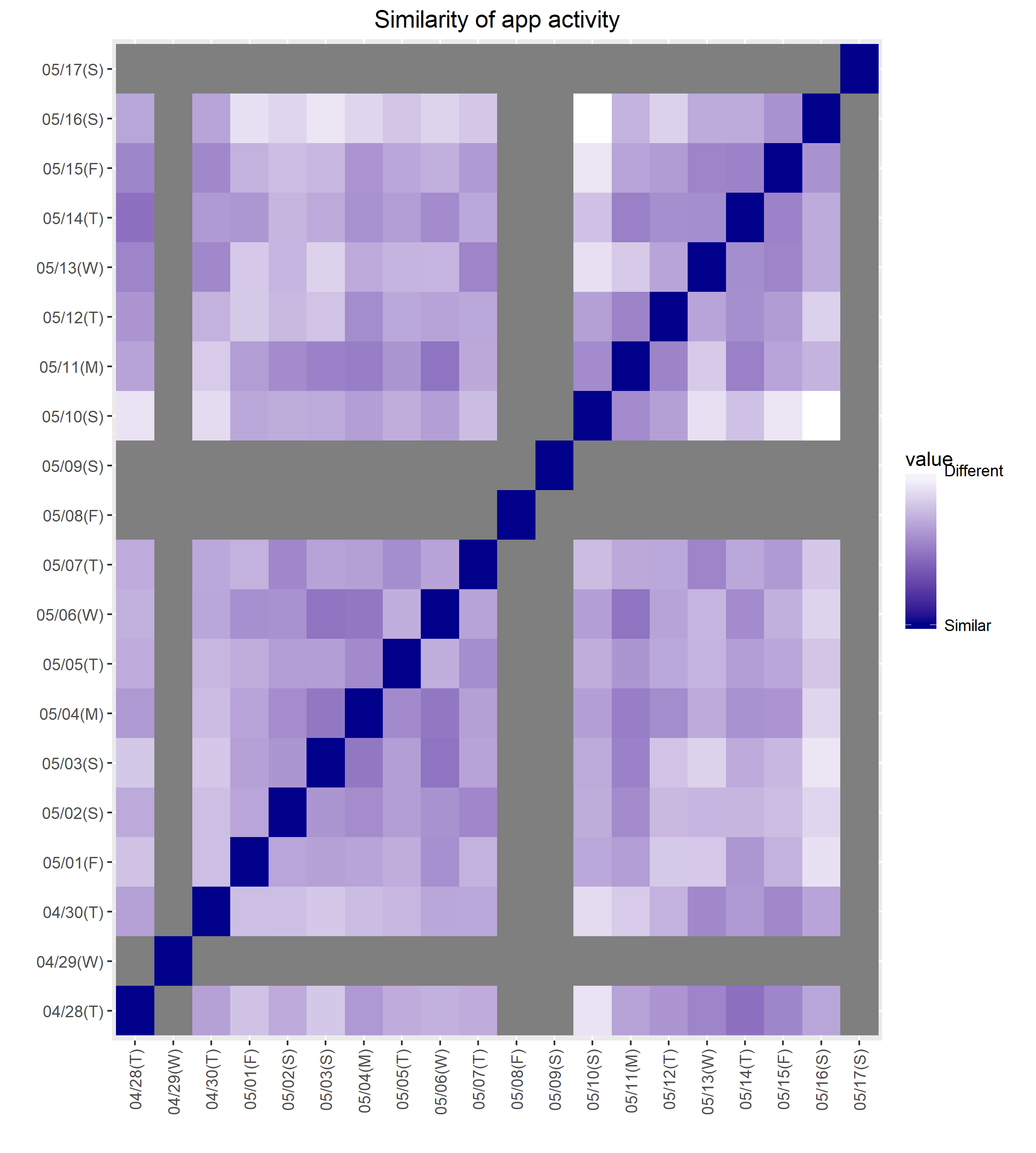}
    \caption{Similarity matrix for valence, arousal, social roles and app usage (columns left to right) of participant 1 before COVID-19, 2 week into state of emergency, and 6 weeks into state of emergency (rows top to bottom)}
    \Description{Similarity matrix showing the similarity between valence and arousal (rows) before, 2 weeks after and 6 weeks after the pandemic declaration.}
     \label{fig:similarity}
\end{figure*}

Figure~\ref{fig:similarity} shows the similarity of valence, arousal, social roles and desktop app usage day to day for Participant 1 including periods of before, 2 weeks into and 6 weeks into the pandemic. The grey cells represent days that we do not have data.
 
Figure~\ref{fig:similarity} shows that for valence, this user reports much more stable valence before the pandemic. Two weeks after the pandemic declaration, the stability of the participant's valence was affected, causing it to change every two to three days. After four weeks, the valence of the participant has seems to have stabilised, showing large portions of time reporting similar distributions of valance each day.

Figure~\ref{fig:similarity} also suggests that the participant's arousal distribution is highly varied during the weekend and early part of the week before the pandemic. During lockdown, the participant's arousal distribution changed across different days, including weekdays and weekends. In the third phase (third row shown in Figure~\ref{fig:similarity}), signifying sixth and seventh week into the lockdown, the arousal seems to stabilise, however, around 11th May, the arousal distribution was highly dissimilar. We found that this was the time when restriction was loosened in Victoria, and up to five visitors were allowed first time indoors and up to 10 outdoors. This potentially explains why the valence on 10th May also shows a different distribution given that the news was released this day.

\begin{figure*}[t]
    \centering
    \includegraphics[width=0.24\linewidth]{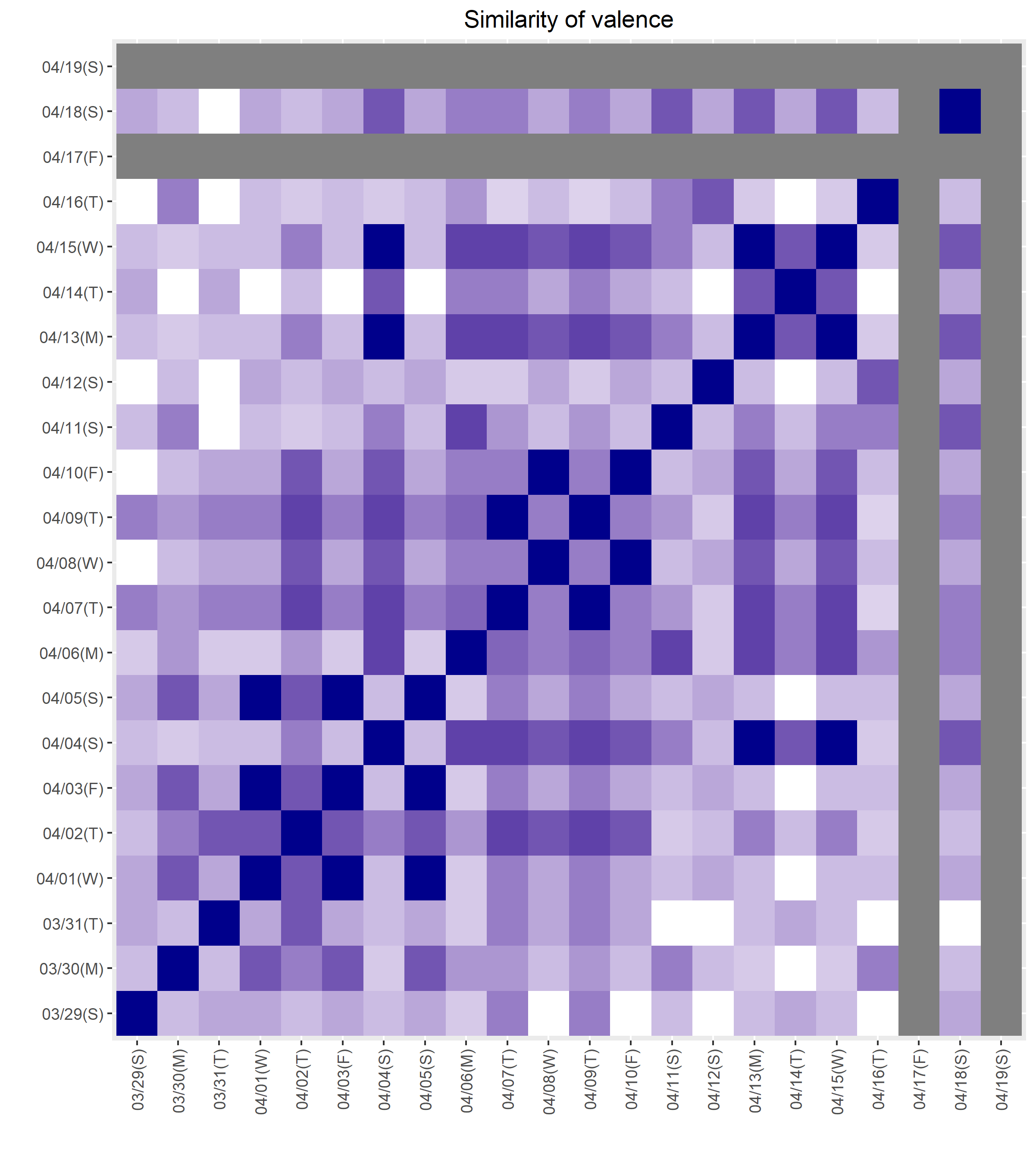} 
    \includegraphics[width=0.24\linewidth]{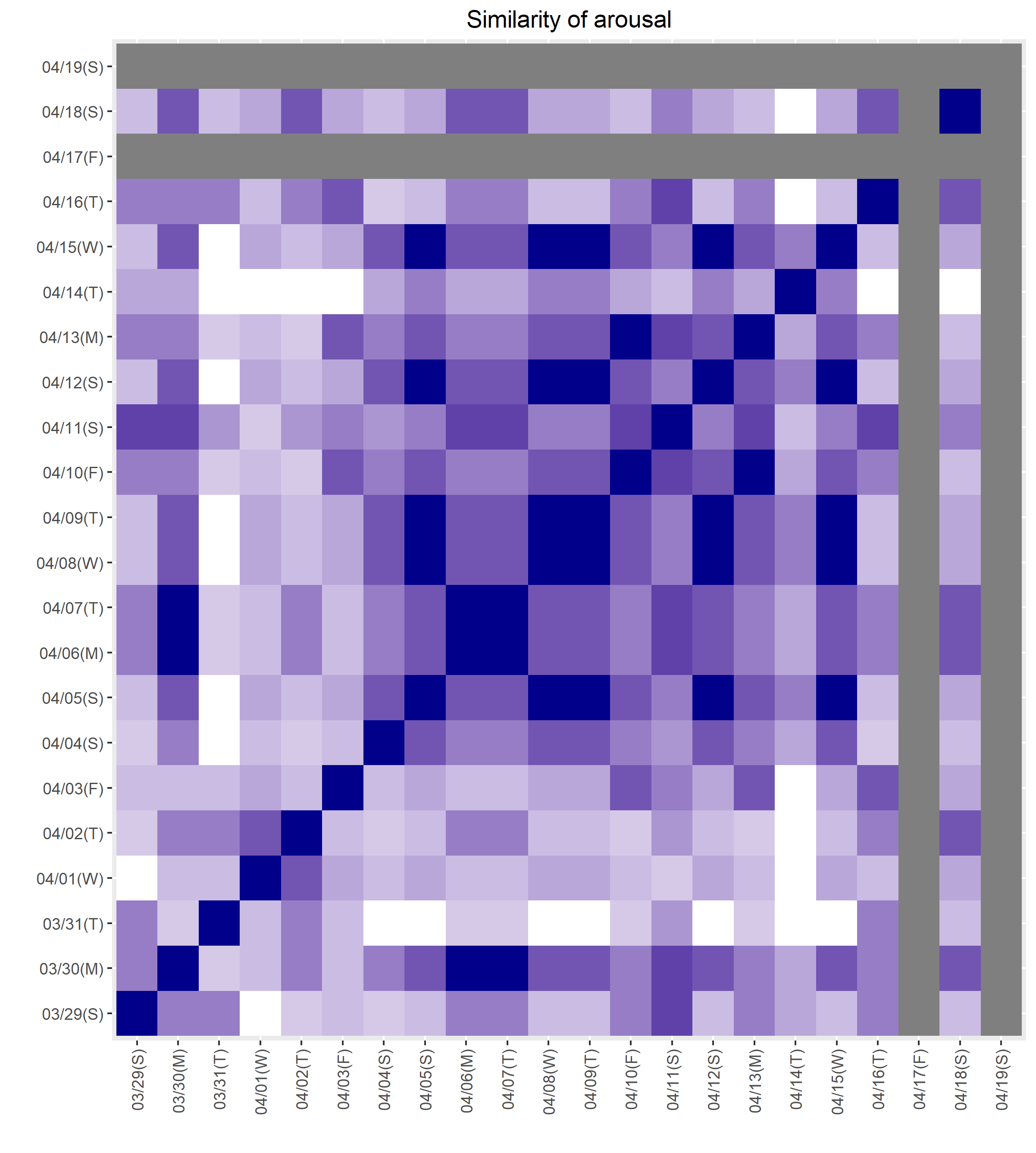}
    \includegraphics[width=0.24\linewidth]{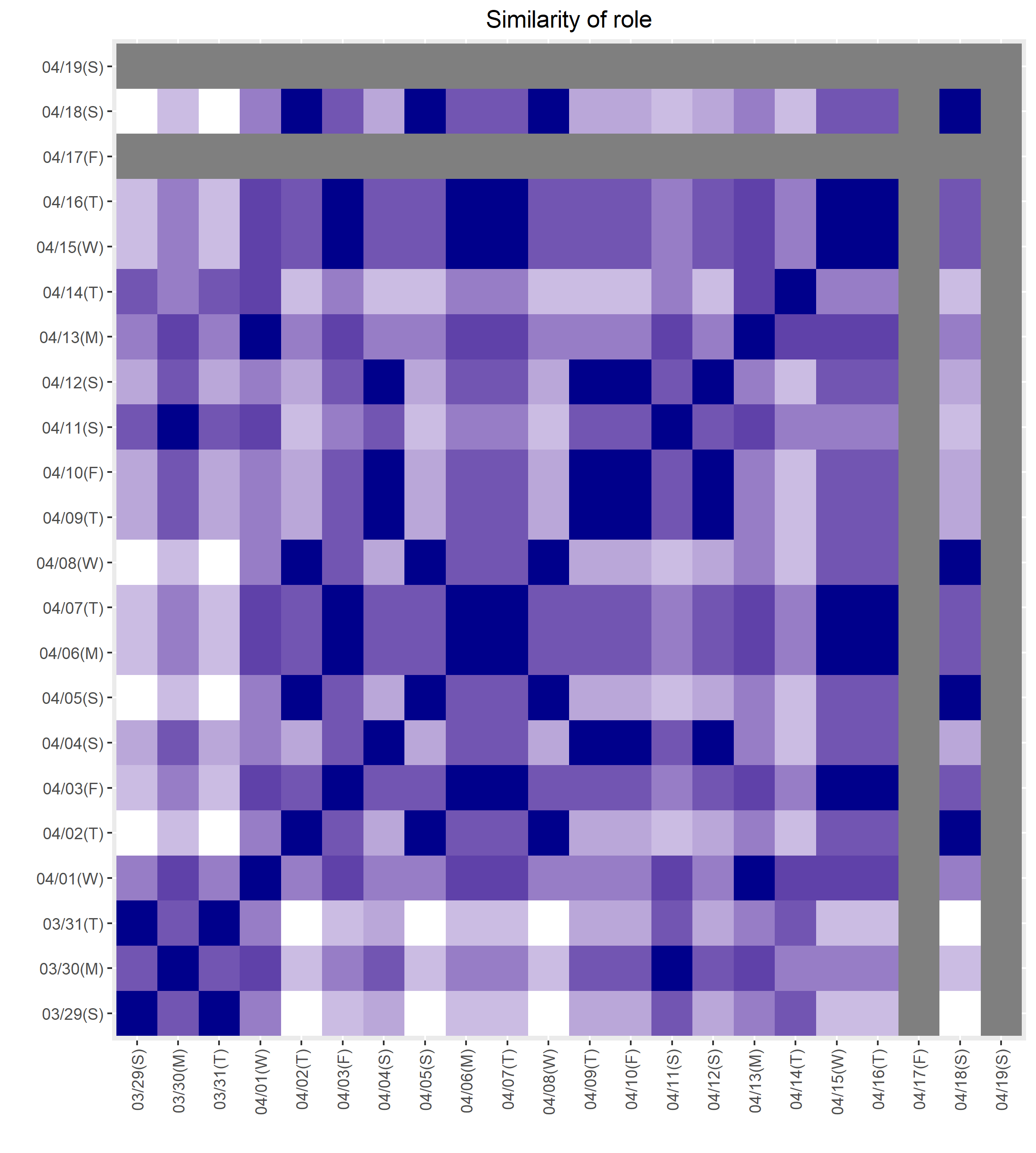}
    \includegraphics[width=0.24\linewidth]{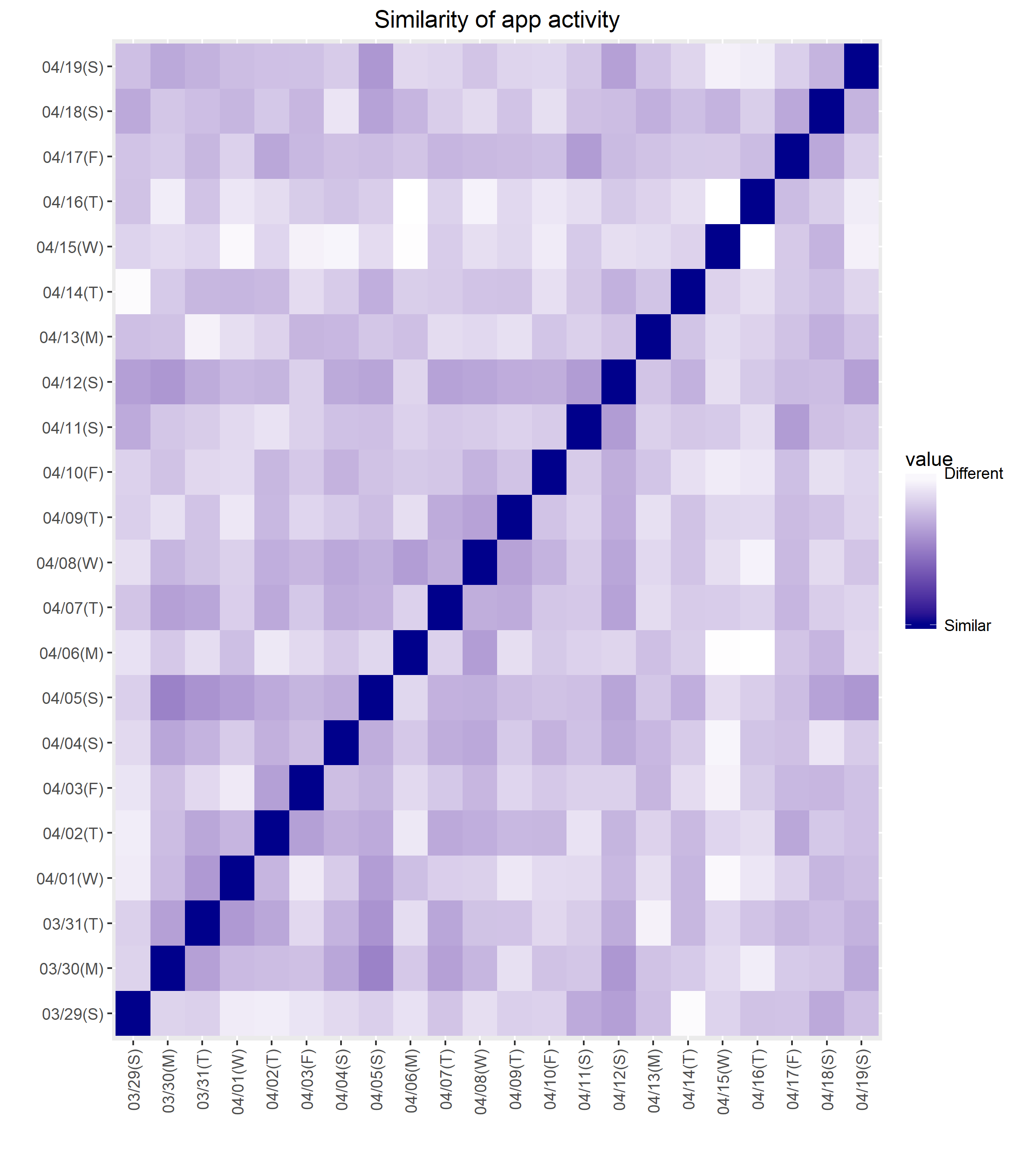}
    
    \includegraphics[width=0.24\linewidth]{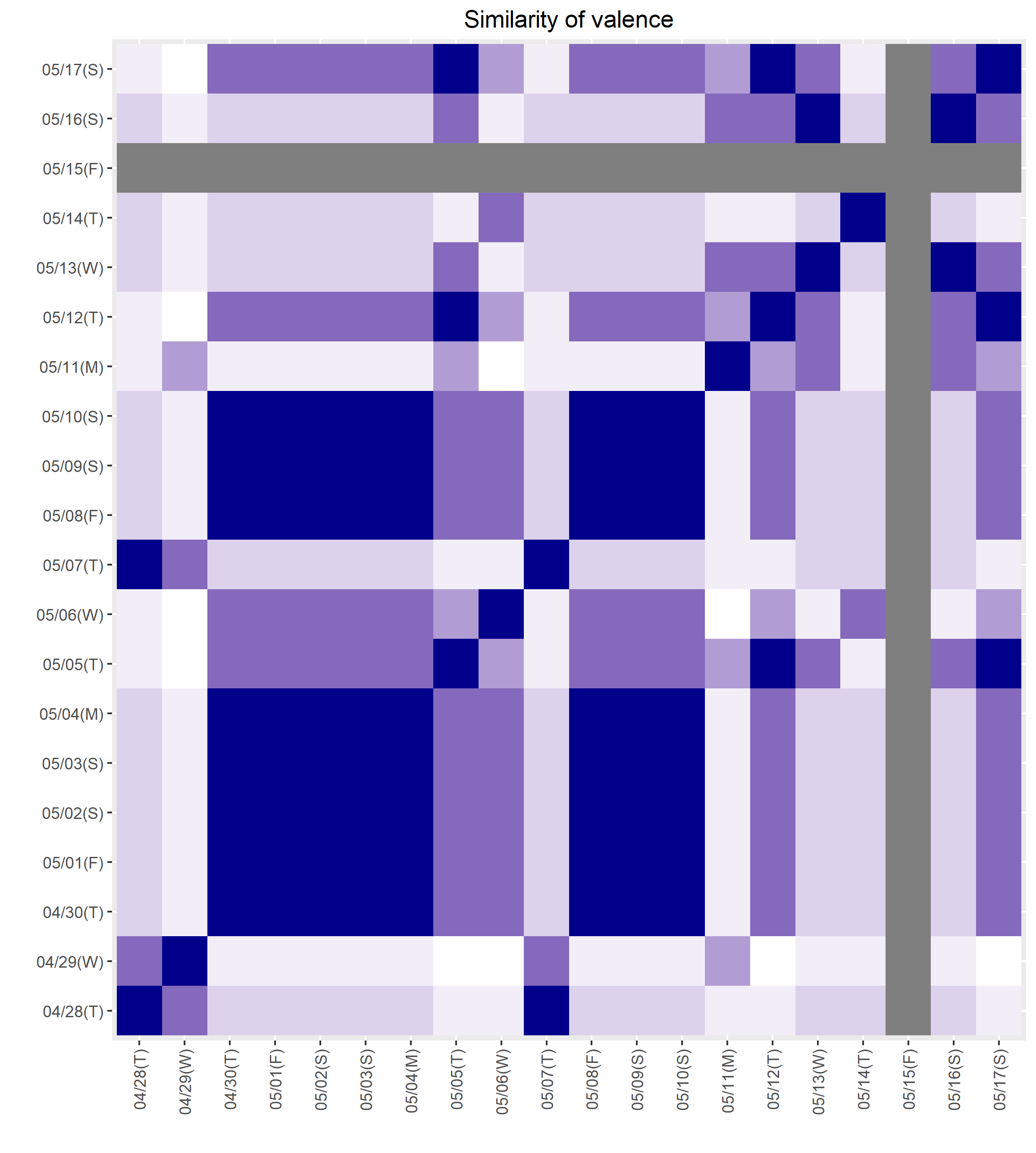}
    \includegraphics[width=0.24\linewidth]{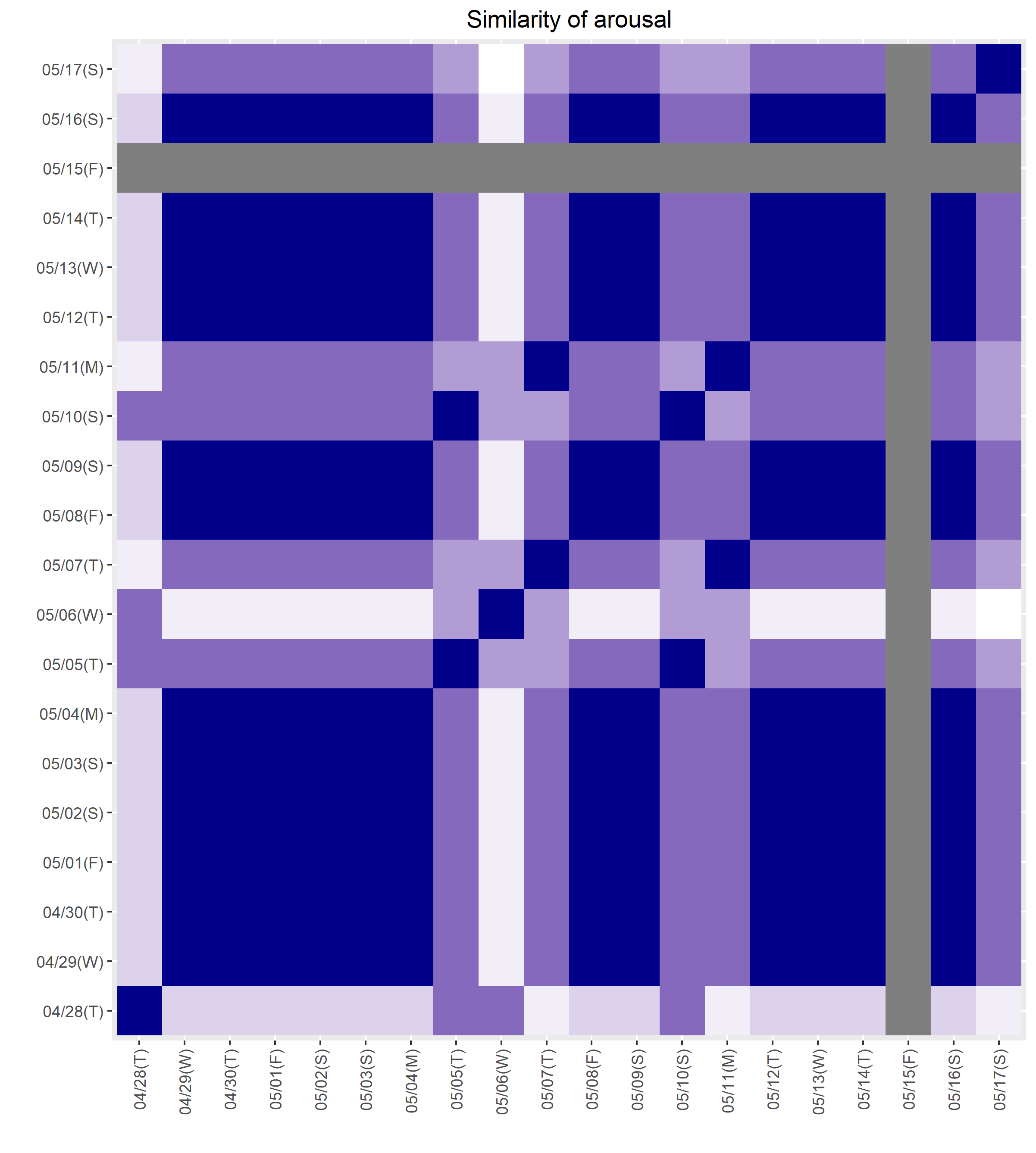}
    \includegraphics[width=0.24\linewidth]{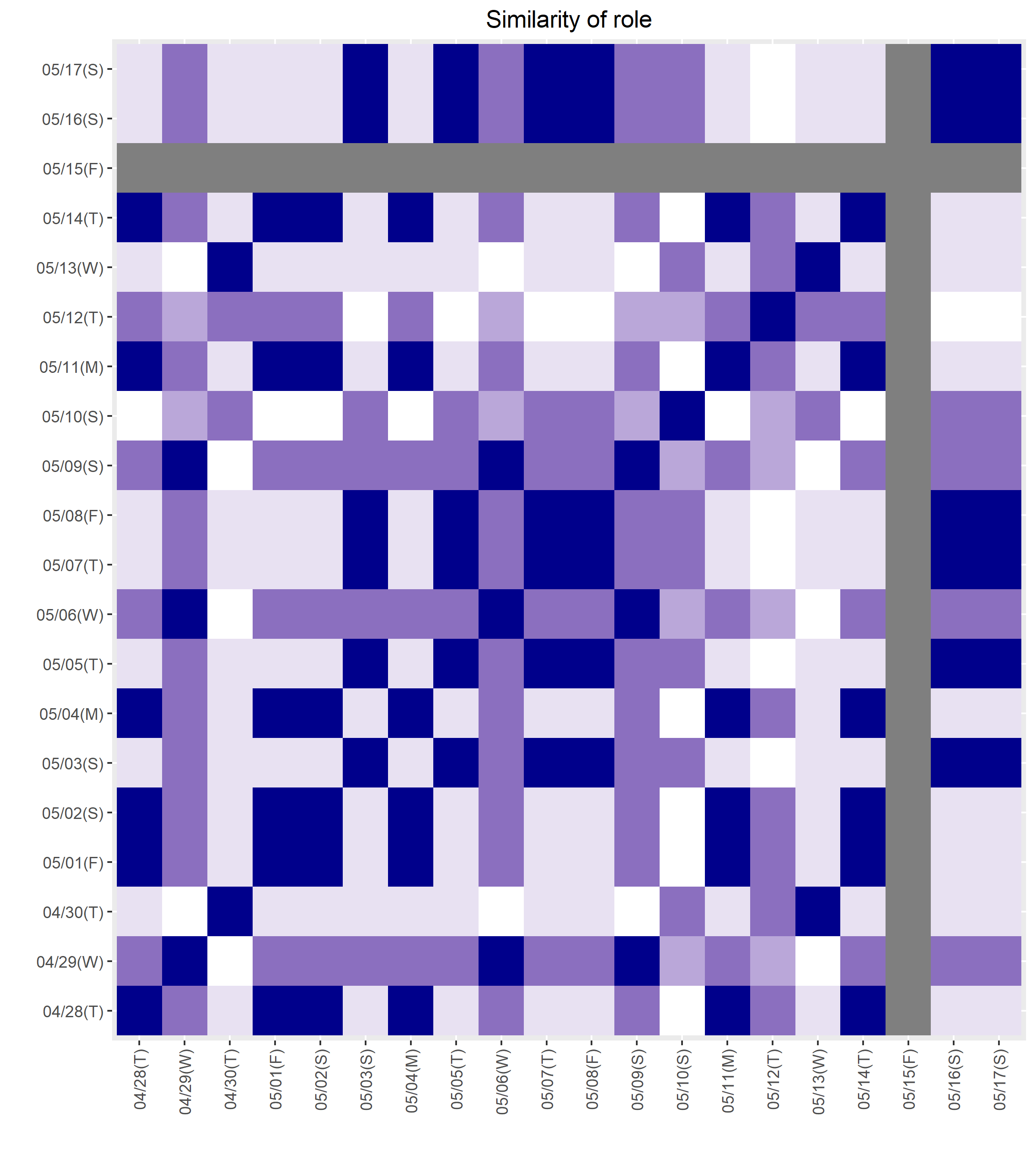}
    \includegraphics[width=0.24\linewidth]{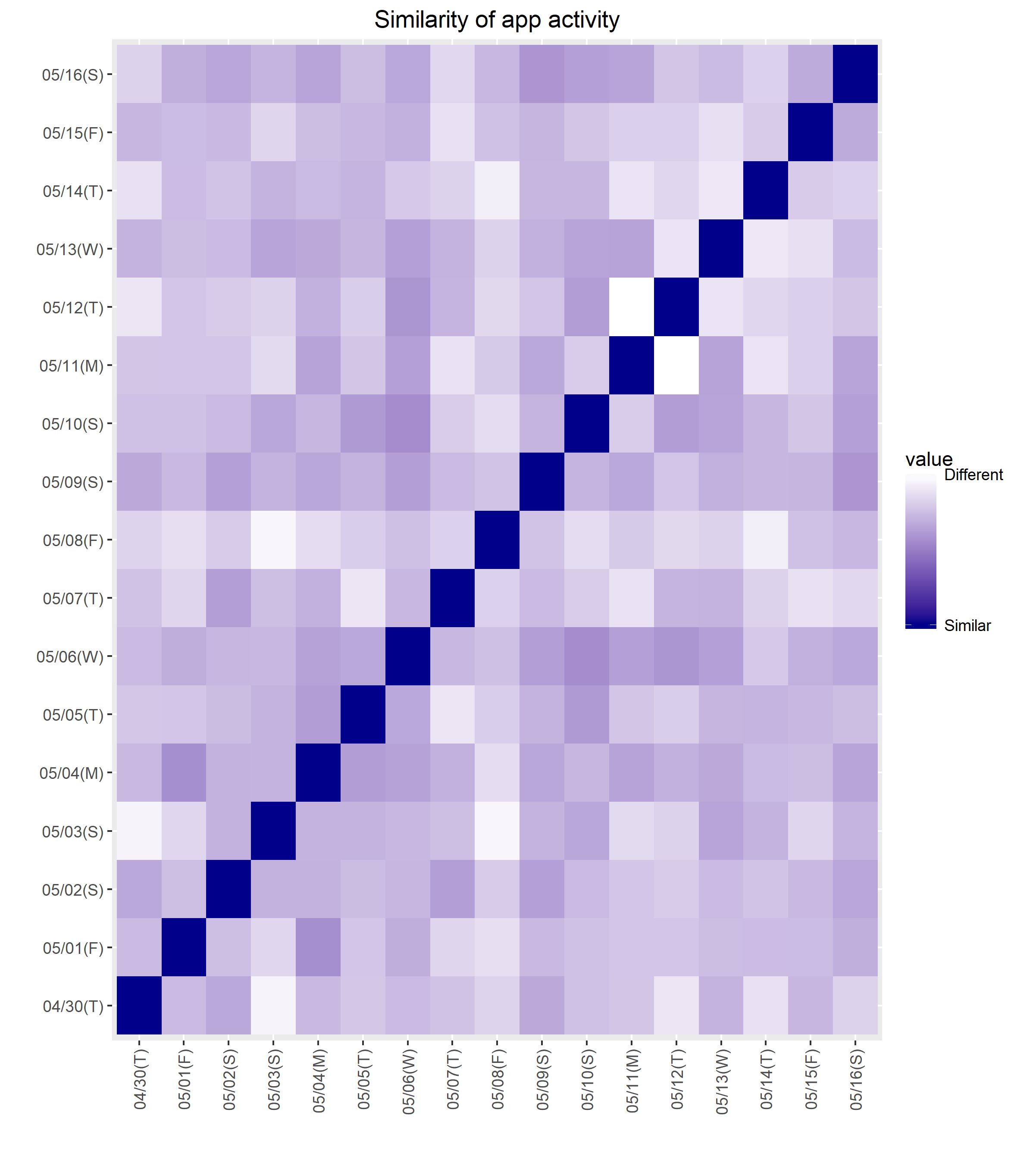}
    \caption{Similarity matrix for valence, arousal, social roles and app usage (columns left to right) of participant 4 for 2 week into state of emergency, and 6 weeks into state of emergency (rows top to bottom)}
    \Description{Similarity matrix for participant 4 valence, arousal, social roles and app usage 2 weeks after and 6 weeks after the pandemic declaration.}
     \label{fig:similaritysam}
\end{figure*}

Interestingly, the social role of the participant shows a large amount of instability at 6 weeks after the pandemic declaration for this participant. This could be caused by changing between work and private roles as they are becoming accommodated to working at home. This is especially interesting considering the charts do not show a large change in stability before the pandemic and 2 weeks in. We can also observe similar behaviour in social role for an another participants in Figure~\ref{fig:similaritysam}. 

Figure~\ref{fig:similaritysam} shows similarity matrices for participant 4's survey responses and the mobile app usage distribution.
The figure suggests that the valence of the individual varied less in the later weeks of the pandemic, along with the arousal. At the start of the pandemic, the valence distribution was found to be different from day to day. Six weeks into the pandemic, the behaviours had stabilised and only changed on the Tuesday and Wednesday and Thursday of each week.

During the pandemic, participant 4 had more varied work and private roles later in the pandemic than earlier in the pandemic.
\section{Discussion and Limitations}\label{dis}
This study found that the impact that the pandemic had on individuals varied.

For only one of three participants, we found that COVID-19 caused more combined work and private roles while working at home. The study also indicates that the days become unstable in terms of social role for some participants during the latter weeks of lockdown. Beyond the scope of this study, this findings shed light on to scheduling and productivity management research. A tool that is capable in assisting individual's work and private job management while working from home can benefit people in being more productive. There are a few works which have been devoted to productivity assistance or inferring concentration level in open work place~\cite{9097829}. In~\cite{liono2020intelligent}, the authors proposed a novel task recognition framework based on app usage for productivity assistant in our daily life.   

Furthermore, only one of the three had a significant impact on happiness. Although the happiness varies in the beginning weeks of the pandemic, the happiness of participants seem to stabilise in the later weeks. This shows that the individuals were not impacted in terms for COVID-19 as much as was hypothesised. Therefore, working from home could be considered a viable alternative for work as we come out of the COVID-19 pandemic.
 
This study is limited in the amount of participants and all from the same occupation and we do not find significant evidence that the COVID-19 lockdown impacted the work and happiness of the majority of participants in this study.
This means that former methods of helping the work of knowledge workers could also be effective during COVID-19. This research however, does not have enough participants to produce any generalisable findings.

This study discovers that during COVID-19 lockdown, all participants are involved in longer work hours. And it is also observed that the distribution of valence, arousal, and the corresponding app usage patterns are much highly varied during the early weeks of lockdown in comparison to the latter weeks. 

As of writing this paper, the pandemic within Australia and the world is far from over. This methodology could therefore be used to measure the impact that COVID-19 has coming out of the pandemic and the future of work. While a larger sample of people participating from a range of different professions and backgrounds is required to measure the true impact of the pandemic, the methodologies developed in this article could help in characterising the behaviour patterns of individuals coming out of the pandemic.

The changing nature of the world we are currently living in is an urge for new ways to think about the work that we do. Further research could look into the effectiveness of working from home for a variety of different occupations. Many novel technologies are yet to come to assist people in continuing work from home as we come out of the pandemic.

Another future direction of this study is predicting application usage and human behaviour in times of crisis, instead of just exploring its effect. This would assist those who are transitioning to work from home. Machine learning models and recommendation systems can be developed further to support this type of research.

\section{Conclusion}\label{conclusion}
This study used a novel dataset to help get an idea of the individual behaviour patterns of people during COVID-19.

This study found that the impact that COVID-19 had on social roles and valence varied between participants. With the majority of participants not showing a significant change in valence nor assumed social role. Furthermore, this study finds that the valence distribution of an individual stabilised over time while the different social roles assumed by the individual increased, showing a resilient adaptation during the lockdown.

We end this paper with an invitation for future research into the effectiveness of more working from home studies during and coming out of COVID-19, looking into improving the productivity and well-being of individuals in these troubling times.

\begin{acks}
The authors would like to acknowledge Shohreh Deldari, who coded the original implementation of IGTS for which we adopted for this study, and who was very helpful in helping us understand the implementation. We would further like to acknowledge the fantastic people working at the front lines trying to fight the COVID-19 virus and keep us safe and healthy.
\end{acks}

\bibliographystyle{ACM-Reference-Format}
\bibliography{bibliography}

\end{document}